\documentclass{preprint}

\usepackage{natbib}%
\usepackage[figuresright]{rotating}

\usepackage{url}
\usepackage{algorithm}

\usepackage{algpseudocode}
\usepackage{tikz}
\usetikzlibrary{arrows,automata,backgrounds,fit,patterns,petri,shadows,shapes}
\usepackage{float}
\usepackage{subcaption}
\usepackage{bm}
\usepackage{amssymb}
\usepackage{mathtools}

\usepackage{amsmath,amsthm}

\newtheorem{proposition}{Proposition}

\theoremstyle{definition}

\theoremstyle{remark}

\jname{International Transactions in Operational Research}
\jvol{XX}
\jyear{20XX}
\doi{xx.xxxx/itor.xxxxx}

\begin{document}

\title{Optimal product launch in uncertain competitive markets: A multi-agent adversarial risk analysis framework}

\author[Arce et al.]{Pablo G. Arce\affmark{a,b,$\ast$},  Sonali Das\affmark{c} and David Ríos Insua\affmark{a}}

\affil{\affmark{a}Inst. Math. Sciences, Spanish Nat. Research Council, C. Nicolás Cabrera, 13, Madrid, 28049, Spain}
\affil{\affmark{b}Universidad Autónoma de Madrid, Escuela de Doctorado, Francisco Tomás y Valiente, 11, Madrid, 28049, Spain}
\affil{\affmark{c}Dept. Business Management, University of Pretoria, Lynnwood Rd, Pretoria, 0002, South Africa}

\thanks{\affmark{$\ast$}Author to whom all correspondence should be addressed (e-mail: pablo.garcia@icmat.es).}

\historydate{Received DD MMMM YYYY; received in revised form DD MMMM YYYY; accepted DD MMMM YYYY}

\begin{abstract}
Product launching is a nuanced problem that entails choosing numerous decision variables, including timing, pricing, quality, and marketing strategy. Furthermore, decision-makers must account for multiple uncertainty sources. These include endogenous factors, such as product quality and the strategic decisions of competitors and consumers; as well as exogenous elements, like market evolution and regulations. This paper proposes a novel general framework to support product launch decisions in the presence of multiple competitors and buyers using adversarial risk analysis, mitigating the strong common-knowledge assumptions typical of standard game-theoretic approaches in this domain. We also introduce variants that account for multiple product purchases and integrate discrete choice models. To overcome the inherent computational intractability of modeling multi-agent strategic uncertainty, we deploy a simulation-optimization approach leveraging Bayesian optimization. Finally, we illustrate our proposed framework through a software launch case study focused on optimal timing, pricing, and quality decisions.
\end{abstract}


\maketitle

\section{Introduction}\label{sec:intro}
 A product’s launch into an intended market is a complex strategic decision that can critically affect both the product’s commercial viability, as well as, in some instances, even the long-term survival of the company. Launch decisions entail accounting for multiple variables including its timing, price, quality, and marketing expenditure, to name but a few, and also depends on the very nature of the product itself, as well as the targeted market segment. In addition, product launch decisions are typically made within a competitive environment characterized by the presence of similar offerings from rival companies. 
     To further complicate matters, a product launch process frequently involves multiple sources of uncertainties, some involving our own product, such as its actual quality; others associated with the decisions of competitors, including their release time; others involving those of the consumers, like their actual preferences; and finally, those concerning market conditions, e.g. new regulations or the future state of the economy.  Given the numerous intervening factors, product launching is a complex business decision for managers and has been the subject of intense modeling work, which we briefly review, emphasizing the uncertainty and strategic decision aspects.

Of the numerous uncertainties involved at the launch of a product, the launch date is of particular importance \citep{hultink1997industrial, harris1967new}. In principle, a company seeks to attain a \textit{first-mover advantage}. However, achieving this depends not only on the strategic intent and quality of their own product, but also on the extent of information available regarding competitors’ products. 
The strategic necessity of anticipating competitive entry, and its subsequent impact on optimal pricing and market share, has long been a focal point in the operations literature, particularly for technology products \citep{mesak1990impact}. While a firm's own strategic intent is an endogenous information, optimally responding to anticipated entry is fraught with exogenous uncertainties, as the nature, release timing, and status of the competitors' products are rarely fully known \citep{POULSEN2007129}. 
It is expected though that a first-mover will gain an initial share of niche market segments, which comes along with the benefits of brand recognition and loyalty. However, it can also result in pitfalls, since any defects remaining in the product at launch time will be capitalized on by competitors \citep{WU2019138}, potentially resulting in future market~shrinkage.

Along with the price quoted, buyers respond to the launch of a product based on  {\em noisy signals}, which includes the quality of the product as advertised \citep{erdem1996decision}. With low-involvement products, such as groceries, a buyer can make various decisions at both category and brand levels \citep{miyazaki2021dynamic}. 
However, with high-involvement products intended to last longer, such as cars, and which are typically more expensive   compared to disposable income, buyers generally have different responses. Regardless, for both low- and high-involvement new products, their uptake also depends on how information about it is transmitted within the customer network \citep{sunar2019optimal}. 
     Branding is another relevant factor: a new product launched by an established brand in a captive market preserves the attention of its prospective buyers who would believe that any eventual defects in the product will be efficiently rectified \citep{wang2025impact, singh2022intertemporal}. 

Concerning strategic aspects, several authors have focused on product launching from a game-theoretic perspective.   In particular, this is the case with approaches based  on classical real options  \citep{dixit1994investment,trigeorgis1996real}.
In so doing, price has been treated as an important feature \citep{LI2019287, LIN20201026, 2023LiHao}, along with quality \citep{JAFARZADEHGHAZI2023486}, and, of course, the price-quality combination \citep{pammer2000forecasting,wang2022product}. 
Focusing on a specific pricing strategy, \cite{lotfi2024skim} developed an optimization model for price-skimming, 
  finding that the effectiveness of this strategy highly depends on the  repeat purchases rate.
Several authors have also addressed the impact of multiple supply-chain echelons when deciding the price of the product at launch \citep{Wei2013, Wang2017, SaneZerang2018,   Taleizadeh2019}.  Other authors have paid attention to the price-advertising pair within oligopolistic competition 
 \citep{golan2000estimating}. For certain new products that may need approval before their launch,  such as a new pharmaceutical drug, \cite{limon2023sequential} provides a holistic examination of adopting a sequential versus a concurrent product development approach by companies, under competition and other uncertainties. Under demand uncertainty for a new product, \cite{ hitsch2006empirical} used dynamic programming to monitor product sales, which can also guide a company to exit a market if sales are low. 
 On the other hand, addressing demand uncertainty, \cite{balter2025innovation} modeled a monopolist decision on the timing, quality, and positioning of a new product relative to an established one, finding that uncertainty often delays the introduction and that such new product may be added alongside, rather than immediately replacing the old one. 
Overall, the conclusion is that the launch of a new niche product by a company in a competitive market is a complex decision, in which a large part of the product's success depends on the `optimal' timing of the launch \citep{hultink1997industrial, savin2005optimal, calantone2007clustering}.
 


 While traditional game-theoretic models have provided valuable insights into competitive equilibrium regarding the aforementioned  price, timing, and quality tradeoffs (see, e.g., \cite{dube2007competitive}), most of the approaches mentioned, 
including those drawing on real options, entail strong common-knowledge assumptions, as discussed in \cite{Hargreaves:2004} or \cite{raiffa}. These assumptions are highly debatable in competitive business domains because they neglect key strategic uncertainties about competitors' and customers' decisions.

Thus, in part to mitigate the above mentioned issue, this paper introduces a broad decision analytic framework to support product launch decisions by a company  based on adversarial risk analysis (ARA).  An in-depth computational and conceptual comparison between game-theoretic and ARA approaches in the security domain is provided in \cite{banks2022}. 
Recent advances have also developed computational ARA methods for multi-agent sequential and security games \citep{EKIN2022}, showing how Bayesian simulation avoids common-knowledge requirements in adversarial environments with continuous decision spaces and private information.
 ARA  business applications are much less frequent with 
  references including \cite{deng2015application} who considered
 pricing strategies in the re-manufacturing domain; 
 \cite{au2021numerical} centered on auctions; 
 \cite{rasines2024personalized} who dealt with personalized pricing 
 strategies targeting a single buyer; and \cite{softearlier} 
   who considered software launch timing decisions with a single buyer. 
   Thus, in this paper
  our main contributions are:
   \begin{itemize}  
   \item A more realistic model for launching a product in which a company simultaneously chooses multiple 
   launch product features in the presence of multiple competitors and a diverse consumer base, providing a novel ARA-based perspective,  mitigating standard common-knowledge conditions  in earlier game-theoretic based approaches including those drawing on real options;
    \item  From the point of view of the ARA methodology, novel structural templates for a decision maker facing multiple heterogeneous competitors and multiple heterogeneous customers are presented;
   \item Besides the more normative expected utility ARA template, a novel combination of ARA with 
   descriptive discrete choice models is introduced;
   \item A combination with stochastic knapsack 
   problems is introduced to handle multiple product purchases;
   \item The required modeling and computational perspectives are illustrated with a realistic software launching use case, reflected through reproducible software.

   \end{itemize}
   
   For this, Section 2 outlines the overall structure of the problem in three stages: It first presents the decision-making problem faced by the supported company; Next, it analyzes the buyers’ behavior, starting with an expected utility model and subsequently integrating it within a discrete choice framework, and differentiating between homogeneous and heterogeneous markets;  Finally, it addresses the competitors’ decision-making problems.
   Section 3  presents modeling and computational challenges through a case study in software launching, focusing on 
  decisions concerning quality, timing, and pricing, with emphasis on strategic considerations and sensitivity analysis to deliver
     managerial insights. Section 4 explores multi-item purchasing decisions embracing stochastic knapsack problem formulations within the general framework.
 We conclude with pointers to future extensions. Appendix A details the parametric settings and data used within the case studies. Proofs of propositions are available in Appendix~B.
 
 Code and full hyperparameter specifications for the algorithms and data 
to reproduce all experiments are made available at \url{https://github.com/pablogarciarce/ProductLaunching}.

\section{General model}\label{general_model}

The scenario of interest involves determining the optimal features of a product to be launched by a company in the presence of several competing companies and multiple potential buyers, with different goals, needs, and constraints. Let ${\bm o}  = (o_1, ..., o_m)$ denote the $m$ features that influence buyers' decisions for such a   product,  assimilated into multiple objectives that they aim to optimize, such as maximizing usability, minimizing price, or maximizing quality. Some of these observed product features may be exact, like its price, while some others are only based on random estimates, like, in general, its quality. 
Assume for now  that each buyer will purchase at most one unit of the product within the corresponding product life-cycle time period $[0,T]$, an assumption revisited in Section 4. 

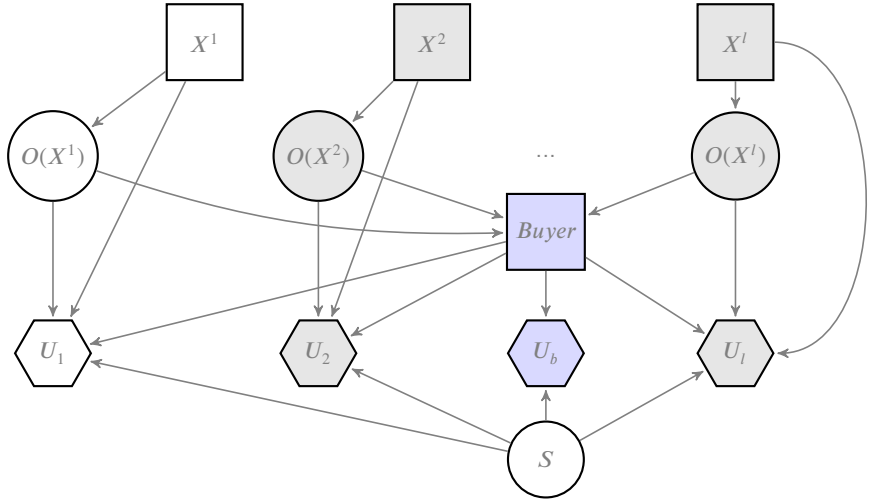
\begin{figure}[htbp]
\centering
\begin{tikzpicture}[->,>=stealth',shorten >=1pt,auto,node distance=1.5cm,semithick]
  \tikzstyle{uncertain}=[circle,pattern=north east lines,
                                    fill=gray!50,
                                    thick,
                                    minimum size=1.0cm,
                                    draw=black]
\tikzstyle{circule}=[circle,thick,minimum size=1.0cm,
                                    draw=black]
\tikzstyle{circule_gray}=[circle,thick,minimum size=1.0cm,fill=gray!20,                             draw=black]
\tikzstyle{circule_gray2}=[circle,thick,minimum size=1.0cm,fill=orange!20,                             draw=black]
  \tikzstyle{attacker_utility}=[regular polygon,regular polygon sides=6,
                                    thick,
                                    minimum size=1.0cm,
                                    draw=black,
                                    fill=gray!20]
  \tikzstyle{attacker_utility2}=[regular polygon,regular polygon sides=6,
                                    thick,
                                    minimum size=1.0cm,
                                    draw=black,
                                    fill=orange!20]
  \tikzstyle{defensor_utility}=[regular polygon,regular polygon sides=6,
                                    thick,
                                    minimum size=1.0cm,
                                    draw=black,
                                    fill=white]
  \tikzstyle{three_dots}=[regular polygon,regular polygon sides=6,
                                    thick,
                                    minimum size=1.0cm,
                                    draw=white,
                                    fill=white]
 \tikzstyle{consumer_utility}=[regular polygon,regular polygon sides=6,
                                    thick,
                                    minimum size=1.0cm,
                                    draw=black,
                                    fill=blue!15]                                   
  \tikzstyle{attacker_decision}=[rectangle,
                                    thick,
                                    minimum size=1cm,
                                    draw=black,
                                    fill=gray!20]
  \tikzstyle{attacker_decision2}=[rectangle,
                                    thick,
                                    minimum size=1cm,
                                    draw=black,
                                    fill=orange!20]
  \tikzstyle{defensor_decision}=[rectangle,
                                    thick,
                                    minimum size=1cm,
                                    draw=black,
                                    fill=white]
  \tikzstyle{consumer_decision}=[rectangle,
                                    thick,
                                    minimum size=1cm,
                                    draw=black,
                                    fill=blue!15]                                  
  \tikzstyle{texto}=[label]
  \node[defensor_decision] (X1) [shift={(-2,0)}] {$X^1$};
  \node[attacker_decision2] (X2) [shift={(1,0)}]{$X^2$};
  \node[attacker_decision] (Xl) [shift={(5,0)}]{$X^l$};
  \node[circule](OX1) [shift={(-4,-1.5)}]{$O(X^1) $};
  \node[circule_gray2](OX2) [shift={(-0.5,-1.5)}]{$O(X^2) $};
  \node[three_dots](...) [shift={(2.5,-1.5)}]{...};
  \node[circule_gray](OXl)   [shift={(5,-1.5)}]{$O(X^l) $};
  \node[consumer_decision] (B) [shift={(2.5,-2.5)}]{$Buyer$};
  
  \node[defensor_utility]  (U1) [below of=B, below of=OX1, yshift=0.4cm] {$U_1$};  
  \node[attacker_utility2]  (U2) [below of=B, below of=OX2, yshift=0.4cm] {$U_2$};
  \node[consumer_utility]  (Ub) [below of=B, yshift=-0.1cm] {$U_b$};
  \node[attacker_utility]  (Ul) [below of=B, below of=OXl, yshift=0.4cm] {$U_l$};
  
    \node[uncertain]  (S) [below of=Ub, yshift=0.2cm] {$S$};
    
  \path (X1) edge    node {} (OX1)
            edge    node {} (U1)
        (X2) edge    node {} (OX2)
            edge    node {} (U2)
        (Xl) edge    node {} (OXl)
            edge[bend left=90]    node {} (Ul)     
        (OX1) edge [bend right=10]    node {} (B)
            edge    node {} (U1)
        (OX2) edge    node {} (B)
            edge    node {} (U2)
        (OXl) edge    node {} (B)
            edge    node {} (Ul)
        (B) edge    node {} (U1)
            edge    node {} (U2)
            edge    node {} (Ub)
            edge    node {} (Ul)
        (S) edge    node {} (U1)
            edge    node {} (U2)
            edge    node {} (Ub)
            edge    node {} (Ul);
\end{tikzpicture}
\hfill 
\caption{Multi-agent influence diagram for product launching problem.}
\label{pagarcia}
\end{figure}

For a company intending to launch a product, decisions involve choosing the values of $d$ decision variables ${\bm x} = (x_1, ..., x_d)$ such as price, ease-of-use, and timing. There will usually be a model
that relates the features ${\bm o}$ with the decisions ${\bm x}$, 
   expressed as  ${\bm o}(\bm{x})$ (e.g., product's ease of use may depend on the design choices and associated development time). Assume there are $l>1$ companies launching competing products within the market of concern to us, each with their corresponding decision variables 
  ${\bm x}^i, i=1, ..., l$. Our goal is to advise the first company in choosing their 
  ${\bm x}^1$ variables. The global problem is sketched through the multi-agent influence diagram \citep{banks2015adversarial} in Figure \ref{pagarcia}.
  Squares, circles, and hexagons respectively represent decisions, uncertainties, and evaluations. Colors reflect distinct 
  agents. 
  Exogenous uncertainties affecting product performance, like 
    macroeconomic conditions, are modeled through a generic  chance node $S$.


  We next sketch the problem faced by the supported company, and, step-by-step, handle the elements that facilitate incorporating the required strategic uncertainties. For each step, we state conditions ensuring existence of the incumbent solution, and discuss the associated computational and modeling~issues.


\paragraph{{\bf Supported company decisions}}  
  Assume that the advised company launching the new product decides on its features ${\bm x}^1$ by maximizing
  its expected utility with respect to its utility function $u_1$ \citep{french}. Suppose for the moment  that we 
    have available the probability $\pi({\bm x}^1)$ of a customer buying the product 
    from the advised company, designed with features ${\bm x}^1$.  Then, if there are  $n$ customers with the same
  purchasing propensity, the purchase of the product can be modeled 
     through a binomial experiment $Bin \big(n, \pi({\bm x}^1) \big)$, assuming, to start with, independent purchasing processes among the $n$ customers, an  assumption later modified.

Let $p_1$ denote the first company's selling price,   
 typically one of the decision variables
 within ${\bm x}^1$, as well as one of the ${\bm o(x}^1)$ product features considered by the buyer. Let $c_1$ 
denote the corresponding production costs. Then, the 
  first company's expected utility associated with its 
  decision ${\bm x}^1$ would be   
\begin{equation}\label{kkbarna}
    \psi({\bm x}^1) = \sum_{j=0}^n 
    \left[ \binom{n}{j} \pi({\bm x}^1)^j (1 - \pi({\bm x}^1))^{n-j} \right] \times u_1 (j \times p_1 - c_1).
\end{equation}
  The company's objective would be to find its optimal decision ${\bm x}^{1}_*$ by maximizing its expected utility $\psi({\bm x}^1)$ subject to constraints ${\bm x}^1\in {\cal  X}^1$ affecting its decision variables, reflecting e.g.\ appropriate price ranges. 
 The compactness  of $ {\cal  X}^1$  and the continuity of $\psi $ with respect to ${\bm x}^1$ would guarantee the existence of an optimal ${\bm x}^{1}_*$. The continuity of the expected utility derives from the continuity of $u_1$ in $p_1$ and of $\pi$ in ${\bm x}^1$, leading to the following result. 
  \begin{proposition}
\label{lemma_1}
If the utility function $u_1$ is continuous in $p_1$ for fixed $j$ and $c_1$, 
the purchasing probability $\pi$ is continuous in ${\bm x}^1$ and 
${\cal X}^1$ is compact, then ${\bm x}_*^1$ exists.
\end{proposition}

  \noindent   From a computational perspective,  solving the resulting expected utility maximization problem 
    (1) requires, in general, stochastic optimization techniques as those described in \cite{POWELL}.
  From a modeling perspective, $u_1$ is standard in Bayesian decision analytic terms \citep{french},
but estimating $\pi({\bm x}^1)$ has a strategic nature. 
   We tend to improve 
 such type of forecasts by considering the problem 
 as in the ARA approach (see \cite{gomez2025forecasting, banks2022} for discussions and experiments) as we now address.

\paragraph{{\bf Buyers' decisions}}  To facilitate estimation of $\pi({\bm x}^1)$, consider the  buyers' decision making problem 
  from the perspective of the supported company. A buyer's decision to purchase from a specific company depends on the observed signals ${\bm o}({\bm x}^i)$ associated with the $i$-th company's product, for $i = 1, 2, .., l$.  Then, if $u_b$ denotes the buyer's utility function, and $p_b(s)$  the density modeling their beliefs about the exogenous factors, the buyer will acquire the product from the first company if they derive the highest expected utility in doing so, expressed through 
\begin{equation}\label{kkoviedo}
    \int u_b({\bm o} ({\bm x}^1), s)\,p_b(s)ds \geq \max_{i \neq 1} ~\int  u_b({\bm o} ({\bm x}^i), s)\,p_b(s)ds .
\end{equation}
 In most cases, only partial information about  $u_b$ and $p_b$ will 
  be available
 to the company. As usual within the ARA framework \citep{banks2015adversarial}, suppose that we model such  partial
 information in a Bayesian manner through a random utility function $U_b$ and a random density $P_b$. Similarly, the ${\bm x}^i$ decisions 
 of the competitors will be modeled as 
random variables ${\bm X}^i$. Without loss of generality, the
random $U_b$, $P_b$, and ${\bm X}^i$ will be defined over 
an appropriate common probability space \citep{Chung}. 
 Then, for any decision ${\bm x}^1$ of the first 
 firm, 
the probability that the buyer selects its product is estimated through   
\begin{equation}
  \pi_{ARA}({\bm x}^1)= Prob \Big(\int U_b({\bm o} ({\bm x}^1), s)P_b(s)ds \geq \max_{i \neq 1} ~\int  U_b({\bm o} ({\bm X}^i), s)P_b(s)ds\Big) .
   \label{eq:pi}
\end{equation}
A direct application
of the dominated convergence theorem \citep{royden2010real} enables proving the existence and continuity 
of the purchase probabilities \eqref{eq:pi}. 
\begin{proposition}
If the utilities $u_b$ in the support of 
$U_b$ are almost surely (a.s.) integrable with respect to the corresponding $p_b$, the probabilities
(\ref{eq:pi}) are well-defined. Moreover, if the functions  
$u_b({\bm o} ({\bm x}^i ), s)$ are a.s. continuous in ${\bm x}^i$,
and there is an integrable function $f(s)$
such that 
$|u_b({\bm o} ({\bm x}^i ), s)| < f(s)$, then    such 
probabilities are continuous in ${\bm x}^1$.
\end{proposition}

\noindent   Computationally, $\pi_{ARA}({\bm x}^1)$ would be estimated by Monte Carlo (MC), by sampling 
from $(U_b, P_b, ({\bm X}^i)_{i=2}^l)$, and consequently estimating whether
or not the expected utility derived from the first company's product is maximal as in (\ref{kkoviedo}).
 If so, we count this as a success, then iterate, and finally estimate the probability through the empirical frequency.

   Observe though that since we are not advising the buyers, just the first producer,
    we could expect them not to behave as \textit{expected utility maximizers}, e.g.\ due to their access only to partial information about products, their cognitive biases, their use of decision heuristics  or implementation errors within their decision-making processes. 
    An alternative approach from the consumer behavior literature \citep{mcfadden1973conditional,train} to estimate buyer choice probabilities 
    uses discrete choice models, leading to random utility 
    models and probabilistic choices, see De Palma et al. (\citeyear{de2008risk}) for 
    a discussion and comparison with expected utility models.    
    A prime example, later illustrated, is the multinomial logit (MNL) version of the Plackett-Luce model \citep{luce59, f5079a1-8ca5-3727-a405-0a82390327b7} which assumes an underlying weighted additive utility function 
    \citep{gonzalez2018utility} $u_b({\bm o} ({\bm x}), s)$,
    with weights $\lambda _j \geq 0, \sum_j \lambda_j=1$,
    with an error 
    model leading to the MNL conditional choice probability
\[ Prob ({\rm buyer \,\, chooses}\,\,{\rm company}\,\,{\rm  1}  | s) =
 \frac{  \exp ( \sum _{j=1}^m \lambda_j o_j (({\bm x}^1),s) ) }
{ \sum_{i=1}^l \exp ( \sum _{j=1}^m \lambda_j o_j (({\bm x}^i),s) ) }. \]
Finally, as in  (\ref{eq:pi}), taking into account the uncertainty about various ingredients 
 (utility weights $\Lambda_j$ and competitors' decisions ${\bm X}^i$)
we would write 
\begin{equation}\label{kkroswell}
\pi_{DC}({\bm x}^1) = \mathbb{E}\left[ 
\frac{  \exp ( \sum _{j=1}^m \Lambda_j o_j (({\bm x}^1),s) ) }
{ \exp ( \sum _{j=1}^m \Lambda_j o_j (({\bm x}^1),s) ) + \sum_{i=2}^l \exp ( \sum _{j=1}^m \Lambda_j o_j (({\bm X}^i),s) ) } \right].  \end{equation}
The existence and continuity of the estimated probabilities in this case is proved under similar conditions to those of Proposition 2.
\subparagraph{{\it The case of heterogeneous buyers}}
In many domains, buyers are most definitely not homogeneous: 
 they belong to different market segments, each with 
 similar features and utility functions, and characterized by a variety of 
 features such as age, location, gender, or socioeconomic status. 
  Suppose there are $K$ market segments, each characterized by a utility function 
  $u_{b_j}$ and a density $p_{b_j}$, with each customer belonging to the $j$-th 
  segment with probability $p_j$, $j=1,2,...,K$.  Repeating procedure \eqref{eq:pi}
  for each segment, we have uncertainty about the beliefs and preferences at each segment,
  which we model with the corresponding random utilities $U_{b_j}$ and probabilities  
   $P_{b_j}$,  and obtain the buying propensity
 $\pi_j({\bm x}^1)$, and conclude that the 
  buying model with $n$ customers would conform to a
  mixed binomial experiment 
    $Bin \big(n, \pi _H ({\bm x}^1)=\sum_{j=1}^K p_j \pi_j ({\bm x}^1) \big)$.
  If the conditions of Proposition 2 hold for each market segment, the global purchase probability is well-defined
and is continuous in ${\bm x}^1$.

 A more detailed approach 
  would consider a hierarchical model based on a buyer type $\tau $ affecting both the utility function  
$u_b({\bm o} ({\bm x}^1), s, \tau )$, and the 
 probability model $p_b(s | \tau )$, with buyer type distributed 
 according to a density $g(\tau )$. \cite{assaf2020bayesian} 
  provide a related discussion within the context of 
  principal-agent problems. Then, a buyer of type
 $\tau $ will acquire the product from the first company if 
\begin{equation*}
    \int u_b({\bm o} ({\bm x}^1), s, \tau)\,p_b(s| \tau)ds \geq \max_{i \neq 1} ~\int  u_b({\bm o} ({\bm x}^i), s, \tau)\,p_b(s| \tau)ds .
\end{equation*}
As before, introducing random variables to model the competitors' decisions 
 (${\bf X}^i$), the buyer's utility ($U_b$)
and the exogenous beliefs ($P_b$)
leads to 
\begin{equation}
  \pi_{\tau } ({\bm x}^1| \tau) = Prob \Big(\int U_b({\bm o} ({\bm x}^1), s, \tau)P_b(s| \tau)ds \geq \max_{i \neq 1} ~\int  U_b({\bm o} ({\bm X}^i), s, \tau)P_b(s| \tau)ds\Big). 
  \label{heter}
\end{equation}
  Then, the global choice probability would be 
\(
\pi_{\tau } ({\bm x}^1 )  = \int   \pi_{\tau } ({\bm x}^1| \tau) g (\tau ) d \tau ,
\)
%
recovering a mixed binomial purchasing model $Bin \big( n, \widehat{ \pi_{\tau }} ({\bm x}^1)\big)$.

\paragraph{{\bf Decision problems of the competing companies}} 
Finally, observe that a random model $X^i$ for the decision
made by each of the other companies are needed in version \eqref{eq:pi} of the problem,
  as well as in its variants \eqref{kkroswell} and \eqref{heter}. 
 For this, 
consider the problem 
for the $i$-th company from an  ARA perspective. 

Such problem would be symmetric to (\ref{kkbarna})
  corresponding to the first company: given their utility 
 $u_i$, unit selling price $p_i$, production costs $c_i$, 
 choice probability $\pi (x^i)$ and constraints ${\cal X}^i$,
  the $i$-th company would seek for their maximum expected utility decision
\begin{equation*}
    {{\bm x}^i}_ *=\arg \max_{{\bm x}^i\in {\cal X}^i }\psi({\bm x}^i ) = \sum_{j=0}^n 
    \left[ \binom{n}{j} \pi({\bm x}^i )^j (1 - \pi({\bm x}^i ))^{n-j} \right] \times u_i (j \times p_i - c_i).
\end{equation*}
As before, the first company will have only partial information about the $i$-th adversary problem.
Taking into account the uncertainty about the involved ingredients, 
  observe that 
\begin{equation*}
  X^i = X^i_* = \arg\max _{x^i \in {\cal X}^i }  \Psi({\bm x}^i ) = \sum_{j=0}^n 
    \left[ \binom{n}{j} \Pi({\bm x}^i )^j (1 - \Pi({\bm x}^i ))^{n-j} \right] \times U_i (j \times P_i - C_i),     
\end{equation*}
where capitals indicate random versions of the corresponding terms (expected utilities $\Psi$, utilities $U$, prices $P$, costs $C$, probabilities $\Pi$).  Combining the arguments in Propositions 1 and 2, we obtain that the (random optimal) decision $X^i_* $  is \textit{a.s.} continuous.


    Computationally, we would sample from the various random ingredients ($\Pi , U_i, P_i, C_i$) and  
solve the optimization problem to obtain a sample from the 
$i$-th company's decisions ${\bm X}^i$, from which we could reconstruct, if necessary, the distribution by MC.
From a modeling perspective,
 first, $P_i$ and $C_i$ respectively model the first company's beliefs about the price $p_i$ and cost $c_i$ of the $i$-th company.
   A heuristic would base it on their own $p_1$ and $c_1$ with some uncertainty around such values, say normal distributions with  respective means $p_1$ and $c_1$ and standard deviations 
 reflecting our uncertainty.
  Next, $U_i$ would model the first company's beliefs about the 
  $i$-th company's preferences.   One could consider a parametric utility function model and introduce a distribution over the parameters; again, as a baseline, we could use $u_1$ and add some uncertainty around its parameters. 
Finally, consider $\Pi({\bm x}^i )$: a simple approach would model it through a beta distribution with parameters ($\alpha({\bm x}^i), \beta({\bm x}^i)$), judgmentally assessed for a few ${\bm x}^i$ values, then interpolated. Note that with the initial assessments, we could obtain $\pi (x^1)$  and  use it as a baseline for $\Pi({\bm x}^i )$; and, then, recalculate
$\pi (x^1)$ and, consequently, obtain $x^{1}_*$ as a robustness  check.

\subsection{An example}

The proposed ARA-based general 
  product launching approach will be specified with an example within the software launching domain that brings additional modeling issues to the fore. Suppose a software company has some flexibility in launching its product within the current product life-cycle
$[0,T]$, which includes development and testing phases, as well as post-release support for software bugs, although the 
  product remains useful
beyond $T$.
For clarity, let us simplify the case by not modeling 
beliefs about potentially relevant exogenous uncertainties $s$.
 Assume 
 price (in the sense that the cheaper the product, the more likely it will be bought), quality (the higher the quality, the more likely its adoption), and timing of launch (the earlier, the better) are key selling attributes in this market. Potential buyers 
   observe price and timing and have access to some proxy of quality (like an estimate of the number of bugs remaining in the software based on ad hoc tests, or by testing $\beta$-versions).  In this case, the decision variables 
    of the $i$-th company will be $(t_i, p_i, q_i)$, with $p_i$ denoting its price; $t_i$, its release time; and $q_i$, its associated quality advertised at launch. 

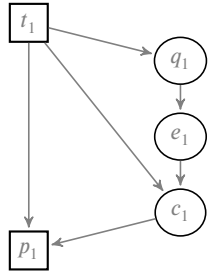
\begin{figure}[htbp]
\centering

\begin{tikzpicture}[->,>=stealth',shorten >=1pt,auto,node distance=1cm,
                    semithick]
  \tikzstyle{uncertain}=[circle,pattern=stripes,
                                    pattern color=gray!40,
                                    thick,
                                    minimum size=1.0cm,
                                    draw=black]
     \tikzstyle{consumer_uncertain}=[circle,
                                    fill=gray!60,
                                    thick,
                                    minimum size=1.0cm,
                                    draw=black]   
       \tikzstyle{competitor_uncertain}=[circle,
                                    fill=gray!20,
                                    thick,
                                    minimum size=1.0cm,
                                    draw=black]                              
\tikzstyle{circule}=[circle,thick,minimum size=1.0cm,
                                    draw=black]
\tikzstyle{elipse}=[thick,minimum size=0.5cm,
                                    draw=black]
  \tikzstyle{utility}=[regular polygon,regular polygon sides=6,
                                    thick,
                                    minimum size=1.0cm,
                                    draw=black,
                                    fill=gray!20]
  \tikzstyle{defensor_utility}=[regular polygon,regular polygon sides=6,
                                    thick,
                                    minimum size=1.0cm,
                                    draw=black,
                                    fill=white]
  \tikzstyle{decision}=[rectangle,
                                    thick,
                                    minimum size=1cm,
                                    draw=black,
                                    fill=gray!20]
  \tikzstyle{defensor_decision}=[rectangle,
                                    thick,
                                    minimum size=0.5cm,
                                    draw=black,
                                    fill=white]
  \tikzstyle{texto}=[label]
  \node[defensor_decision]  (t1) [shift={(0,3)}] {$t_1$};
  \node[defensor_decision]  (p1t1) [shift={(0,0)}] {$p_1$};
  \node[ellipse,draw = black](c1t1) [shift={(2,0.5)}]  {$c_1$};
  \node [ellipse,draw = black] (e1t1) [shift={(2,1.5)}] {$e_1$};
  \node [ellipse,draw = black] (q1t1) [shift={(2,2.5)}] {$q_1$};
    \path (t1) edge    node {} (p1t1)
          (t1) edge    node {} (q1t1)
          (c1t1) edge node{} (p1t1)
          (q1t1) edge node{} (e1t1)
          (e1t1) edge  node{} (c1t1)
          (t1) edge  node{} (c1t1);;
\end{tikzpicture}
\caption{Influence diagram showing relevant software release variables and their relations.}
\label{kk}
\end{figure}

   Concerning constraints, the supported company needs to decide on $t_1 \in [0,T]$ 
  (identifying the need to release within the current product life-cycle)
 and $p_1 \in [a, b]$ (indicating reasonable lower and upper bounds for the price). 
  Note that the price $p_1$ is partly related to the release time $t_1$. 
   On the one hand, the longer the testing time, 
  the higher the development costs; on the other hand, as a result, the smaller the number of bugs left, and hence the lower post-release costs. Pricing should balance both costs.  Moreover, a smaller number of bugs, due to a longer testing time, would entail an increase in quality. At a given time $t_1$, suppose we have a (probabilistic) prediction of the number $e(t_1)$ of remaining bugs, related to the quality assessment $q_1(t_1)$. This facilitates building an estimate of the post-release software maintenance cost and the total cost $c_1(t_1)$ used, in turn, to decide the price $p_1(t_1)$. Figure \ref{kk} displays  the inter-relations between the decision variables $t_1$ and $p_1$, and the relevant product attributes: 
   $q_1(t_1)$, $e(t_1)$, and $c_1(t_1)$ depend on $t_1$ (the ${\bm o} ({\bm x}^1)$ in the notation from  Section 2 displayed in Figure \ref{pagarcia}), whereas 
   $p_1(t_1)$ is based on both $t_1$ and $c_1(t_1)$.

\noindent
For the subsequent narration, remove the notational dependence on $t_1$ and refer to the time, price, and quality of the software launched by the supported company as $(t_1, p_1, q_1)$, and the corresponding number of bugs and cost as $(e_1, c_1)$. 
For the other companies within this market,
 let us characterize their attributes as
  $(t_i, p_i, q_i, e_i, c_i)$, for $i=2,...,l$. Following the arguments in Section 2, 
  we lack access to the exact values of $(t_i, p_i, q_i)$, and assume random distributions over them, denoted as 
 $(T_i, P_i, Q_i)$, $i=2,...,l$.

  Suppose that buyers belong to only one market segment,  
  and let their utility function be $u_b$, representing the archetypal preferences within such a segment. Based on
  \eqref{kkoviedo}, 
they  will choose the first product~if 
\begin{equation*}
     u_b(t_1, p_1, q_1) \geq \max_{i \neq 1} \,\,\, u_b(t_i, p_i, q_i).
\end{equation*}
 Since we do not fully know the utility function $u_b$, 
we introduce a random utility function $U_b$. Then, given a choice of 
 $t_1$ and $p_1$, we estimate the probability that a buyer 
acquires the product from the first company as    
\(
    \pi(t_1, p_1 ) = Prob \big( U_b(t_1, p_1, q_1) \geq \max_{i\neq 1}  U_b(T_i, P_i, Q_i) \big).
\)

For $n$ buyers in the segment, the purchase of the first software is modeled through a binomial distribution
 $Bin(n, \pi(t_1, p_1))$, as before. 
Finally, if $u_1$ is the  utility of the supported company,
the expected utility achieved through the decision $(t_1,p_1)$ 
 would be 
\begin{equation}\label{kkbaiuca}
    \psi(t_1, p_1) = \sum_{j=0}^n  \left[ \binom{n}{j} \pi(t_1,p_1)^j (1-\pi(t_1,p_1))^{n-j} \right] \times u_1\big(j \times  p_1  - c_1  \big).
\end{equation}
The optimal decision is reached by finding the values that provide the maximum expected utility, that is  
\begin{equation}
\max_{t_1\in [0,T], p_1\in [a, b]} \,\, \psi(t_1, p_1).
\label{max_software}
\end{equation} 

\section{A case study in software launching}\label{sec:software}


Let us demonstrate the proposed framework through a case study in software launching covering its entailed modeling and computational issues.
  It is widely recognized that software at the time of launch can be susceptible to failures which can lead to potential and far-reaching consequences
\citep{couce}.  Hence, software companies 
 undertake testing processes to ascertain its 
   quality and reliability before release. Therefore, developing optimal policies and strategies 
   for software launch is a crucial business decision, 
  characterized as a two-level process 
 (software-producers competition for buying consumers)  with multiple 
decision-makers with 
adversarial interests and multiple objectives (see \cite{ruggeri2018decision} for 
  a broad overview 
    focused on deciding the launching time).
 For our demonstration, we adapt the framework outlined in Section 2 to support 
a software development company in choosing multiple release features, in the presence of multiple competitors and multiple buyers, thereby greatly extending the work in \cite{softearlier} who 
 focused on the standard software engineering problem of deciding 
 how long to test before releasing a product, in the presence of just one buyer.  
 As in the basic model from Section 2, we 
 consider a scenario in which customers only buy one product within the product development cycle, as in, e.g., universities investing in just one 
 ERP system for, say, the next five years. This is  
   revisited in Section 4  where multiple product purchases are considered. 
 
\subsection{The case}\label{subsec:example}

 Our case draws on 
  the example in \cite{softearlier} with the actual specific parametric setup and data used described in Appendix A.
 Assume the total number of software faults is fixed, but unknown, and consider a perfect debugging process, thus not
  introducing new bugs. 
   Moreover, assume that fault arrivals can be described through the same process during the debugging and operational phases, once the software has been released.
 Specifically, suppose that the number $e_1(t)$
of bugs in the software known by time $t$ follows a non-homogeneous Poisson process (NHPP) 
  \cite{BASP}. 
   Given failure time data and a prior distribution over the parameters, we obtain a posterior predictive distribution 
  for the number
of failures to be discovered by time $T$ with a Markov chain 
Monte Carlo (MCMC) approach as Appendix A describes.

The cost functions adopted include terms referring to the
testing time undertaken, and to removing bugs
before and after releasing the software, being similar to the classical ones in \cite{Okumoto1979}, and \cite{Zeephongsekul1995},
 although we employ the number $e_1(t)$, rather than the expected number $m_1(t)= E[e_1(t)]$, of
 faults used by those authors. To wit, 
  if $c_1(t)$ designates the total costs incurred by the 
    company when releasing the software at time $t$, we employ  
\begin{equation}\label{cost-game}
c_1 (t) = c_{11} t + c_{21} e_1(t) + c_{31} \left[e_1(T)-e_1(t)\right], 
\end{equation} 
where $c_{11}$ is the testing cost  per unit time until release; $c_{21}$, the cost of removing a fault during testing;  and $c_{31}$, the cost of removing it after release. Typically, fixing an error is more expensive after release so
  that $c_{31} > c_{21}$. Other terms could be added to \eqref{cost-game} to model additional company expenses, such as staff costs, but we adopt this simpler version for illustrative purposes.  
Observe that $ e_1(T)-~e_1 (t)= q_1 (t)$ represents the 
estimated number of bugs that remain in the software after release, and can be viewed as a proxy for software
  quality.
Therefore, by deciding the release time $t$, we obtain the quality proxy $q_1(t)$ and an estimate $c_1(t)$ of the cost, which helps 
 to establish a reasonable range for the price $p_1$. 

  As an initial benchmark, later generalized in Section 3.2, suppose that the decisions of the other companies are modeled as uniform within appropriate ranges, 
 essentially based on past behavior. Thus, for the moment,
  we pay no attention to strategic aspects concerning 
   competitors: these represent what are commonly referred to as level-0 thinkers in the sense of \cite{stahl},  
    and the advised company is treated as a level-1 thinker 
 with respect to its competitors. 
  However, the company will be treated as a level-2 thinker in relation to  consumers who, in principle, aim to maximize expected utility. Specifically, assume the buyers' utility function has the parametric form 
\begin{equation}
    u_b(t,p,q | w_1, w_2, w_3, \rho )= 1-\exp (-\rho \times 
 (- w_1 t - w_2 p - w_3 q ) ),
 \label{eq:utility}
\end{equation}
with non-negative weights $w_1, w_2, w_3$ such that $w_1 + w_2 + w_3=1$.  
 The variables $t$, $p$, and $q$ are normalized so that the utility is not dependent on the 
  natural units chosen (such as year, euros, and number of bugs left); the weights place the objectives on a common scale, while $\rho$ is a risk aversion coefficient corresponding to a constant absolute risk averse (CARA) multi-objective utility function, see  \cite{gonzalez2018utility} for details.
  
  Rather than employing the simulation-optimization approach entailed by 
  \eqref{eq:pi} to assess the purchase probabilities
  $\pi(t_1,p_1)$, let us use an MNL discrete choice model,
    as in \eqref{kkroswell},
\begin{equation*}
    \widehat{\pi }^{MNL}( (t_1, p_1) |w_1,w_2,w_3, \rho) = \frac{\exp{ ( u_b(t_1,p_1,q_1  | w_1, w_2, w_3 , \rho) )      }}{\sum_{i =1}^l \exp{  ( u_b(t_i  ,p_i  ,q_i   | w_1, w_2, w_3, \rho )  )            }}.
\end{equation*}
We next integrate out the uncertainty about
the parameters $(w_1,w_2,w_3) $ and $\rho$, modeled with distributions $\pi(w_1, w_2, w_3)$ and $\pi(\rho)$, 
so that the choice probability for the first product 
with features $(t_1,p_1)$~is 
\begin{equation*}
    \widehat{\pi }^{MNL} (t_1,p_1)  =  \iiiint \widehat{\pi }^{MNL} ((t_1,p_1) |w_1,w_2,w_3, \rho ) \pi (w_1,w_2,w_3) \pi (\rho )
   \,d\rho \,d w_1 \,d w_2 \,d w_3.
\end{equation*}
Once with an estimate of the choice probability $\widehat{\pi }^{MNL} (t_1,p_1)$,
we solve (\ref{max_software}).

  Algorithm \ref{alg:generate_sample}
  estimates by MC 
  the expected utility \eqref{kkbaiuca} associated with decision $(t_1,p_1)$, as well as its expected profit (corresponding to a risk neutral company), with MC sample size $M$.
  In it, we assume available (random) predictive models for $T_j, P_j, E_j(t)$ for 
   two competitors designated $j=2,3$, as well as a predictive model 
    for $e_1(t)$, exemplified 
    in the experiment. 
 This routine would feed a stochastic optimization algorithm 
 to determine the optimal $t_1^*$ and $p_1^*$. 
 Note that the algorithm is highly parallelizable.


\begin{algorithm}
\caption{Expected utility and expected profit of decision $t_1, p_1$} 
\label{alg:generate_sample}
\textbf{Input:} $M$, $n$, $t_1 $, $p_1 $, $c_{11}$, $c_{21}$, $c_{31}$, $e_1(t)$, $T_j$, $P_j$, $E_j(t)$, $j=2,3$  \\
\textbf{Output:} $util$, $profit$
\begin{algorithmic}[1]
\State $\pi = 0$
\State $cost = 0$
\For{$ i \in \{1, \ldots , M \}$}
\State Generate $t_j \sim T_j, p_j \sim P_j, j=2,3$ 
\State Generate $q_j \sim e_j (T) - e_j (t_j) , j=2,3$ 
\State Generate $w_1, w_2 , w_3 \sim \pi(w_1, w_2 , w_3), \rho \sim \pi(\rho)$ 
\State Generate $e_1 \sim e_1(t_1)$ 
\State Compute $q_1=e_1(T) - e_1(t_1) $ 
\State Compute $u_j= 1-\exp (-\rho (-w_1 t_j -w_2 p_j - w_3 q_j)),
j=1, 2,3$
\State Compute $u_j =u_j - u_1, j=2,3$
\State $\pi = \pi  + 1 / (1 + \exp (u_2) + \exp (u_3)) $
\State Compute $c_1 = c_{11} t_1 + c_{21} e_1 + c_{31} q_1$
\State $cost = cost + c_1$
\EndFor
\State $\pi = \pi / M$
\State $cost = cost / M$
\State $util=0$
\State $profit=0$
\For{ $k  = 0, \ldots, n$ }
\State  $ util = util + \binom{n}{k} \pi^k (1-\pi ) ^{n-k} 
u_1 (k\times p_1 - cost) $
\State  $ profit = profit + \binom{n}{k} \pi^k (1-\pi ) ^{n-k} 
 (k\times p_1 - cost) $
\EndFor
\State \textbf{Return:}  $util$, $profit$
\end{algorithmic}
\end{algorithm}

%
\paragraph{{\bf Experiment}}
We use the parametric setting and data in 
Appendix A to exemplify the case study, though 
 it is later thoroughly checked through sensitivity analysis. 
In particular, we adopt a power-law NHPP process ($i.e.$ with mean function $m(t)=at^c$) and gamma priors over its parameters $a$ and $c$ to predict the bugs of the software of the first company, and versions with higher variance for the other two companies. Initially we use a risk-neutral utility function (identity). Samples from the relevant posterior distributions are obtained through a Hamiltonian MC method \citep{nuts} 
  as implemented within the Python package $pymc$, used to predict the number of bugs. 

{\bf Optimization.} Algorithm \ref{alg:generate_sample} estimates the expected utility for each decision $t_1$ and $p_1$.
  In this experiment, the probability models $T_j$, $P_j$, 
 $e_j(t)$, $\pi(w_1, w_2 , w_3)$ and $\pi(\rho)$ do not depend on 
 the decision variables $(t_1, p_1)$ for $j=2, 3$ (Appendix A). Hence we  
   run steps 4-6 in Algorithm \ref{alg:generate_sample} only once, for all combinations of $(t_1, p_1)$.
  
A fundamental computational challenge in multi-agent adversarial risk models is the curse of dimensionality. As the number of strategic competitors and the dimensionality of the decision space increase, exact analytical evaluation of the expected utility surface becomes computationally intractable. To overcome this bottleneck and efficiently navigate the high-dimensional search space, the expected utility maximization problem (7) is handled through Bayesian optimization (BO) \citep{gramacy}. We implement this using the $gp\_minimize$ function from $skopt$, which leverages Gaussian process approximations to construct a surrogate model of the objective function, drastically reducing the number of expensive Monte Carlo calls to Algorithm 1. 
It specifically employs a \textit{Matern kernel}, providing the flexibility to model a wide variety of complex, non-linear objective functions by tuning its hyperparameters through marginal-likelihood maximization. We perform the optimization allowing for up to 200 evaluations of the objective function (the negative expected utility, as we seek to maximize it). This search yields an optimal decision of $(t_1^* =356, p_1^* =8162)$, resulting in an expected profit of $2,840,987$.

\begin{figure}[htbp]
    \centering 
    \hfill
    \begin{subfigure}{0.42\textwidth}
        \centering
        \includegraphics[width=\textwidth]{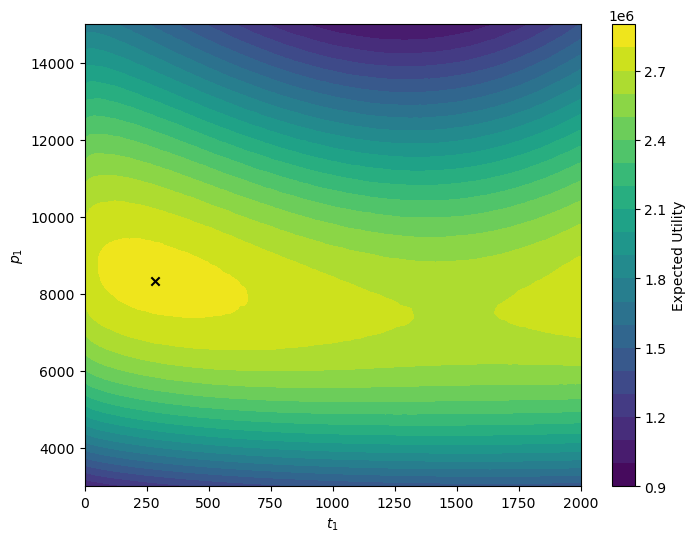}
        \caption{Expected utility with respect to $t_1$ and $p_1$. Optimal decision marked with \textbf{x}.}
        \label{fig:isolines}
    \end{subfigure}
    \hfill
    \begin{subfigure}{0.42\textwidth}
        \centering
        \includegraphics[width=\textwidth]{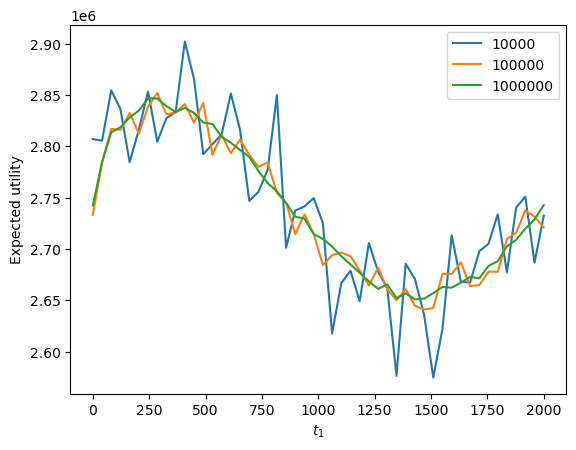}
        \caption{Expected utility for optimal price $p_1^*=8333$ for three MC sizes and various release times. }
        \label{fig:p1_15000}
    \end{subfigure}
    \hfill
    \caption{Expected utility exploration within the experiment.}
    \label{fig:utility}
\end{figure}

As mentioned, optimization can be accomplished through various methods. 
Table \ref{tab:optimizacion_transposed} displays  
results attained with three of them (BO, simulated annealing, brute force) and estimated computation time on a single core (although all methods are amenable to different degrees of parallelization). 
\begin{table}[h]
\centering
\begin{tabular}{c|ccc|c}
  & \(p_1^*\) & \(t_1^*\) & Exp.Prof. & Comp. (s) \\
\hline
Bayesian optimization & 8162 & 356 & 2,840,987 & 55\\
Simulated annealing   & 7711 & 759 & 2,782,931 & 32\\
Brute force           & 8333 & 283 & 2,841,380  & 7202\\
\end{tabular}
\caption{Optimal solutions with three optimization methods.}
\label{tab:optimizacion_transposed}
\end{table}
The setup for simulated annealing is described in
Appendix A.  Brute force is also used
to explore the expected utility surface.
 Specifically, we employ a grid of $100 \times 100$ points over the feasible set $t_1\in[0, 2000]$ and $p_1\in[3000, 15000]$, and an MC sample size of $10^6$. Figure \ref{fig:isolines} represents the estimated expected utility for different $t_1$ and $p_1$ values ($t_1$ in $x$-axis, $p_1$ in $y$-axis), and color-coded 
 expected utility as on its right-hand side. 
  The optimal 
decision would be $(t_1^* =283, p_1^* =8333)$
yielding an expected profit of $2,841,380$, close to the BO solution in value terms. Figure \ref{fig:p1_15000} represents the estimated expected utility as a function of $t_1$ for the 
identified optimal price $p_1^*$, for MC sample sizes 
 $10^4$, $10^5$, and $10^6$, suggesting a relatively flat optimal expected utility surface and comparatively high variance for smaller MC sample sizes. An MC size of $10^6$ is fixed for the rest of the exploration.

\begin{figure}[htbp]
    \centering 
    \hfill
    \begin{subfigure}{0.42\textwidth}
        \centering
        \includegraphics[width=\textwidth]{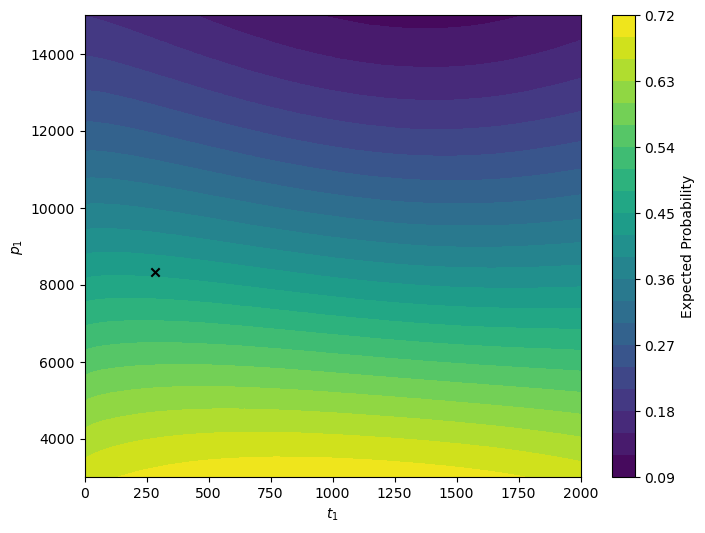}
        \caption{Expected purchase probability with respect to $t_1$ and $p_1$. Optimal decision marked with \textbf{x}.}
        \label{fig:probs}
    \end{subfigure}
    \hfill
    \begin{subfigure}{0.42\textwidth}
        \centering
        \includegraphics[width=\textwidth]{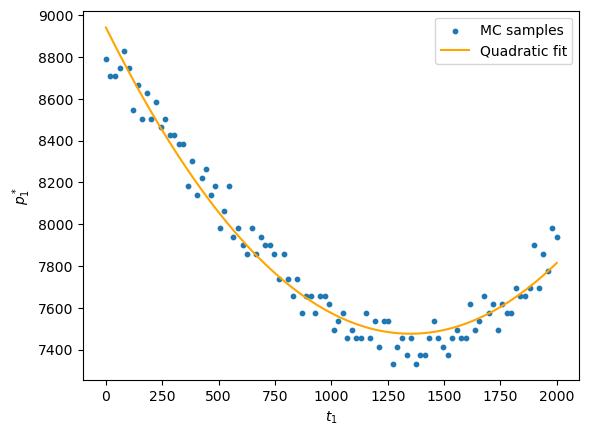}
        \caption{Optimal price $p_1$ as function of release time $t_1$.}
        \label{fig:precio_optimo}
    \end{subfigure}
    \hfill
    \caption{Sensitivity analysis within the experiment.}
    \label{fig:price_probability}
\end{figure}

Figure \ref{fig:probs} represents the
expected purchase probability  $\pi(t_1, p_1)$ from the first company 
given $t_1$ and $p_1$,  with marked probability 0.44 associated 
  with 
the above optimal solution $(t_1^*, p_1^*)$.  Observe that should we aim 
a certain market share, we could identify the corresponding  time-price pairs in Figure \ref{fig:probs}
and use it as constraints within the optimization problem. 
In turn,
Figure \ref{fig:precio_optimo} displays 
the optimal price $p_1^* (t_1)$ given the 
release time $t_1$. Recall that 
such a price would be established as a function of 
the release time. Hence, this curve provides  
  a contingency plan in the sense that we might actually have 
to release before or after the optimal $t_1^*$ and, consequently,
modify the price as provided by this curve.
 Observe that the quadratic fit does not show high variation with $t_1$ within the interval $p_1\in[7200, 9000]$ in this case.

\subparagraph{{\bf Strategic robustness and tipping points}} 
To extract actionable managerial insights and validate the stability of our proposed approach, we systematically stress-test the optimal strategy against variations in key parameters. Because our framework efficiently maps the high-dimensional expected utility surface, we can identify critical {\em tipping points} where a company must fundamentally alter its launch strategy. 

\begin{figure}[h]
    \centering
    \includegraphics[width=\linewidth]{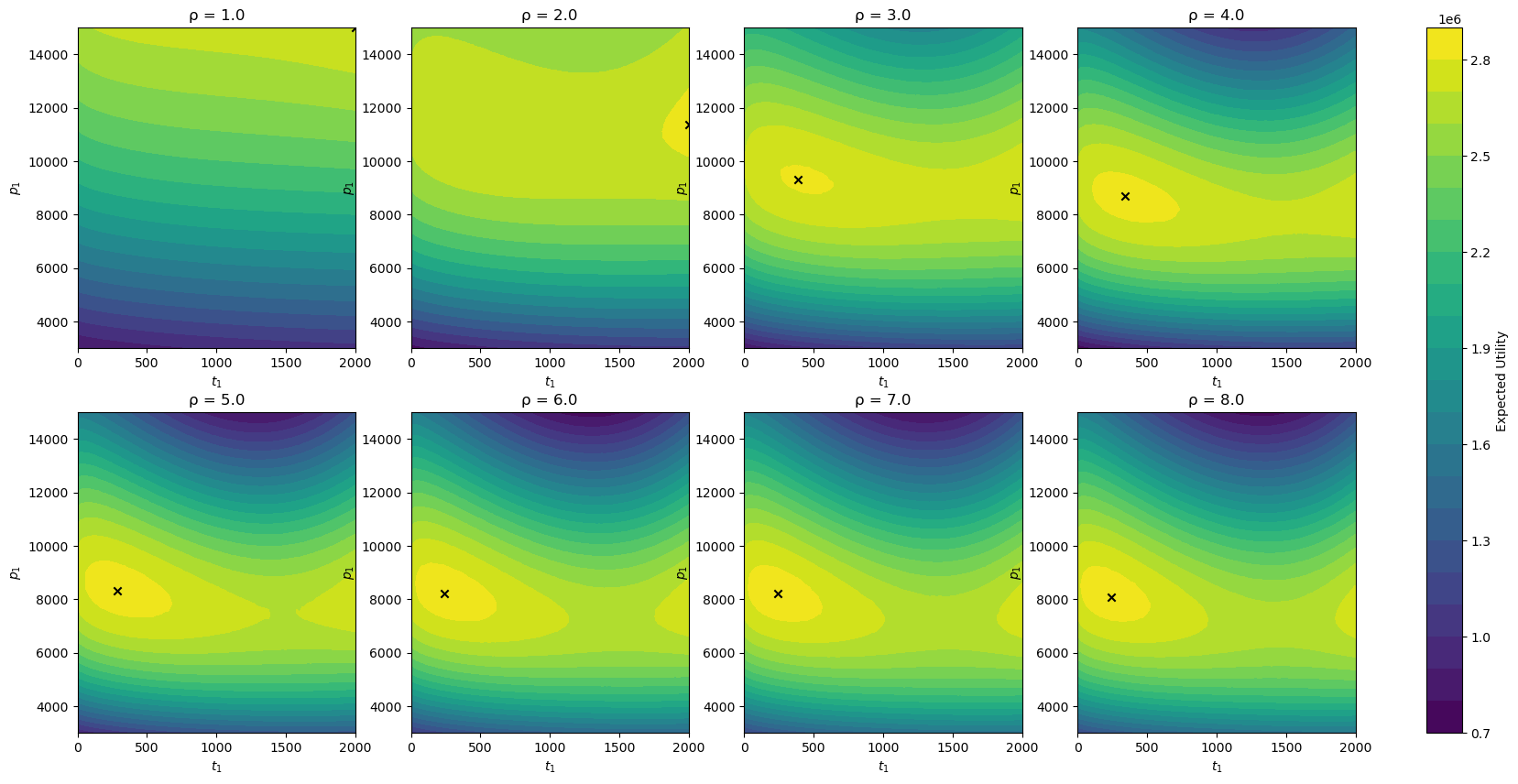}
    \caption{Expected utility depending on $\rho$. Optimal decision marked with \textbf{x}.}
    \label{fig:buyer_risk}
\end{figure}

First, examine the optimal response to shifts in the consumer risk aversion coefficient
$\rho$ 
 in \eqref{eq:utility}, representing e.g. different demographic groups. 
  Table \ref{tab:buyer_risk} presents the 
 optimal results (price, time, expected profit, expected purchase probability) for several $\rho $ values, whereas 
  Figure \ref{fig:buyer_risk} 
  represents the expected utility surfaces associated to these values.
\begin{table}[h]
    \centering
    \begin{tabular}{c|cccc}
        $\rho$ & $p_1$ & $t_1$ & Exp. Utility & Exp. Prob \\
        \hline
        1  & 15000  & 2000  & 3,601,065  & 0.28 \\
        2  & 11364  & 2000  & 2,879,161  & 0.31 \\
        3  & 9303   & 384   & 2,809,326  & 0.39 \\
        4  & 8697   & 343   & 2,832,774  & 0.42 \\
      {\bf   5}  & {\bf 8333}   & {\bf 283}   & {\bf 2,841,380}  &   {\bf  0.44} \\
        6  & 8212   & 242   & 2,846,952  & 0.45 \\
        7  & 8212   & 242   & 2,846,556  & 0.45 \\
        8  & 8091   & 242   & 2,844,337  & 0.46 \\
    \end{tabular}
    \caption{Optimal values depending on risk aversion coefficient.
     Benchmark value $\rho =5$.}
    \label{tab:buyer_risk}
\end{table}
 Observe how, initially, the optimal value corresponds to that maximizing the price and reducing the cost, as the probability changes just slightly. 
  As risk aversion increases, the bigger utility associated to launching earlier and cheaper gains relevance. Thus, 
the optimal price $p_1^*$ tends to stabilize in the 8000–8500 range, whereas the optimal release date $t_1^*$ remains relatively low (below 300). This suggests that beyond a certain risk aversion level, the utility surface does not present significant changes, as  Figure \ref{fig:buyer_risk} shows.

We next perform sensitivity analysis with respect to $c_{31}$ in \eqref{cost-game} within the initial risk-neutral setup.    This parameter is the cost of removing a fault after release, with benchmark $c_{31}=5000$, mapping the framework to different industries where faults are more expensive. 
 \begin{figure}[htbp]
    \centering 
    \hfill
    \begin{subfigure}{0.42\textwidth}
        \centering
        \includegraphics[width=\textwidth]{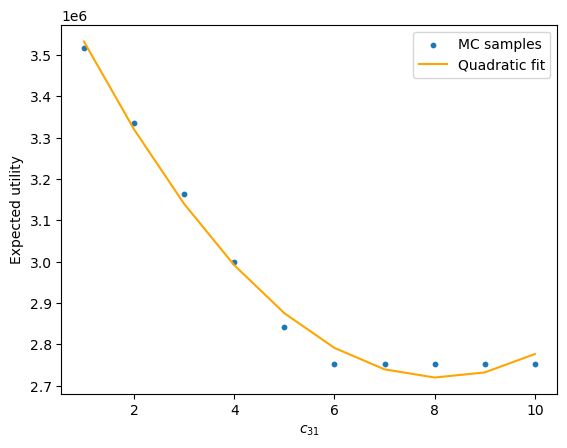}
        \caption{Expected profit as function of $c_{31}$.}
        \label{fig:c31u}
    \end{subfigure}
    \hfill
    \begin{subfigure}{0.42\textwidth}
        \centering
        \includegraphics[width=\textwidth]{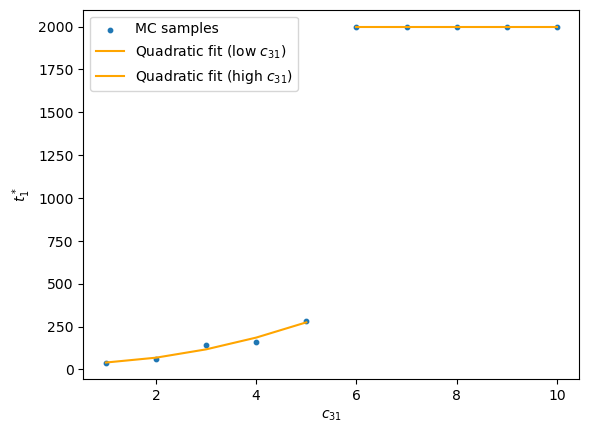}
        \caption{Optimal release time $t_1$ as function of $c_{31}$.}
        \label{fig:c31t1}
    \end{subfigure}
    \hfill
    \caption{Expected profit and optimal time against cost $c_{31}$ (in units of $10^3$). }
    \label{fig:c31s}
\end{figure} 
 Figure \ref{fig:c31s} represents the expected utility and 
   the optimal release time as $c_{31}$ varies. Naturally, the expected utility gets reduced as the after-release cost increases. 
   More interestingly, the framework identifies a critical strategic tipping point regarding  release time. As  Figure \ref{fig:c31t1} illustrates, the relationship is not merely linear: once the post-release penalty ($c_{31}$) crosses a specific threshold, the optimal strategy undergoes a severe regime change. The company abandons early market entry entirely, deferring the launch to the latest possible date ($t_1^* = 2000$) to guarantee negligible post-release bug costs. In Appendix~C, Proposition~\ref{prop:tipping_point} characterizes this discontinuous regime shift, deriving the threshold $\bar{c}_{31}$ where bug-remediation savings overcome the demand penalty of launch delay.
   
\begin{figure}
       \centering
       \includegraphics[width=1\linewidth]{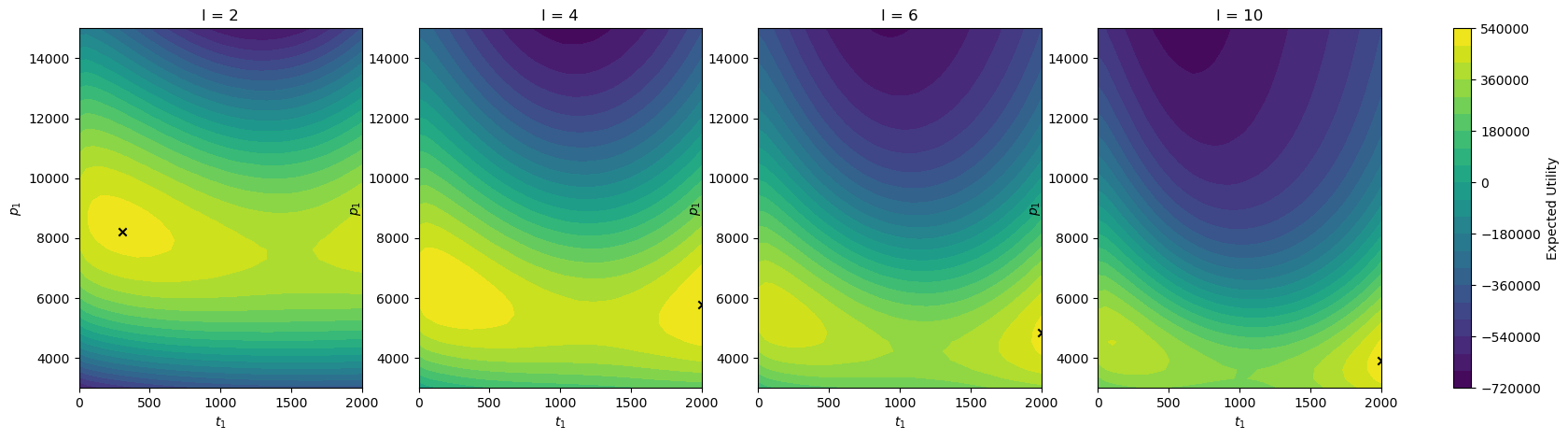}
       \caption{Expected utility depending on the number of adversaries. Optimal decision marked with \textbf{x}.}
       \label{fig:num_adv}
   \end{figure}

   Finally, we perform a sensitivity analysis with respect to the number of competitors, initially fixed at~2. Figure \ref{fig:num_adv} shows the expected utility surfaces for different values of $l$. Observe that, as the number of competitors increases, the location of the optimal solution shifts noticeably. For low number of competitors, the optimal solution favors early release times and higher prices. However, as such number grows, the competitive pressure forces the supported company to move towards later releases and lower prices, seeking to capture market share with a higher-quality product. In addition, and importantly, expected utility surfaces become steeper in these regions, indicating higher sensitivity.

\subsection{Strategic modeling of competitors}\label{kk:sidecars}

 Section \ref{subsec:example} modeled competing companies as
  level-0 thinkers through uniform distributions, overlooking the adversarial aspects of this
   part of the problem, beyond what may be deduced from previous interactions with competitors.
   Such analysis will serve to benchmark these strategic models and showcase
    how a more detailed modeling of competitors may significantly affect the launching strategy. 
   
\begin{figure}[htbp]
    \centering 
    \begin{subfigure}{0.32\textwidth}
        \centering
        \includegraphics[width=\linewidth]{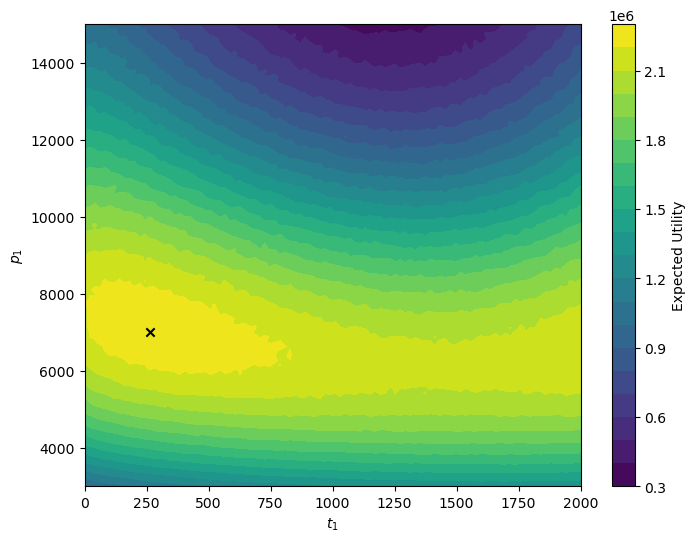}
        \caption{Two standard competitors.}
        \label{fig:utility_intelligent}
    \end{subfigure}
    \hfill
    \begin{subfigure}{0.32\textwidth}
        \centering
        \includegraphics[width=\linewidth]{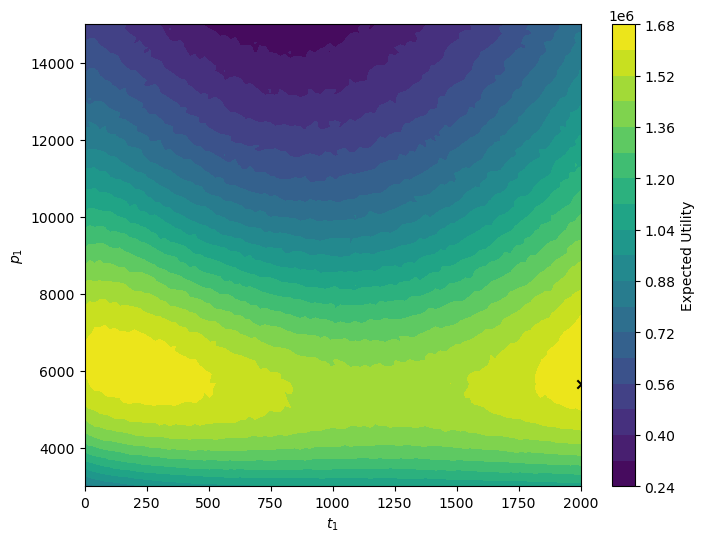}
        \caption{One aggressive competitor.}
        \label{fig:utility_cheapearly}
    \end{subfigure}
    \hfill
    \begin{subfigure}{0.32\textwidth}
        \centering
        \includegraphics[width=\linewidth]{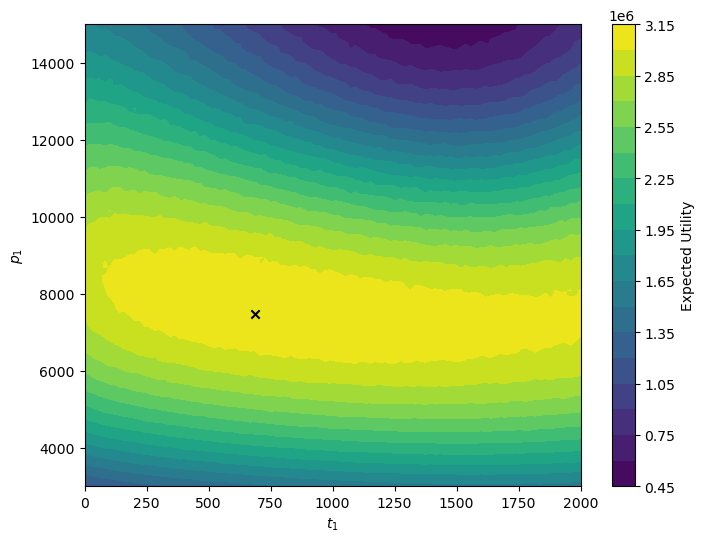}
        \caption{One careful competitor.}
        \label{fig:utility_lateexpensive}
    \end{subfigure}
    \caption{Expected utility against different competitors. Optimal decision marked with \textbf{x}.}
    \label{fig:intelligent}
\end{figure}

\begin{figure}[htbp]
    \centering 
    \begin{subfigure}{0.32\textwidth}
        \centering
        \includegraphics[width=\linewidth]{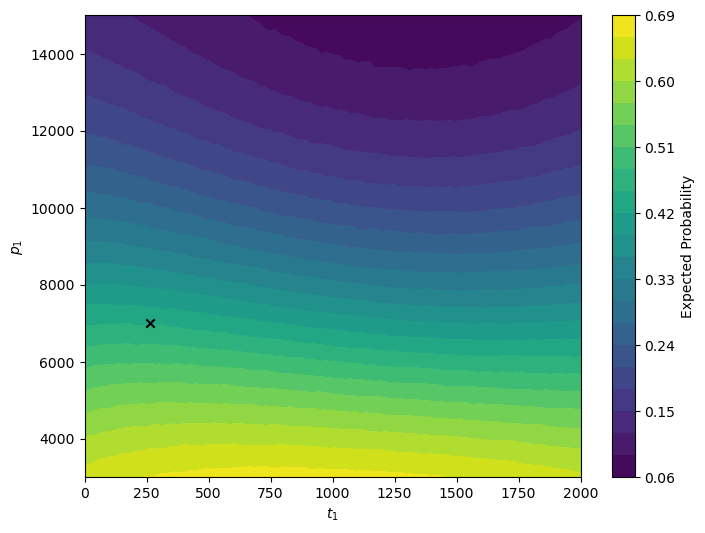}
        \caption{Two standard competitors.}
        \label{fig:probs_intelligent}
    \end{subfigure}
    \hfill
    \begin{subfigure}{0.32\textwidth}
        \centering
        \includegraphics[width=\linewidth]{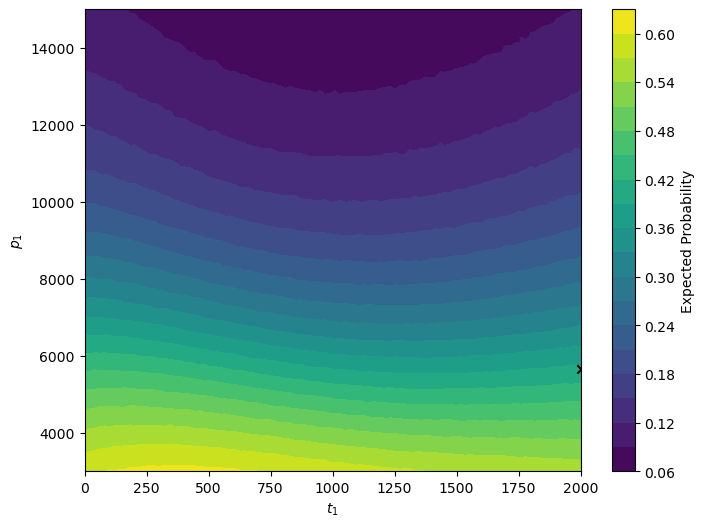}
        \caption{One aggressive competitor.}
        \label{fig:probs_cheapearly}
    \end{subfigure}
    \hfill
    \begin{subfigure}{0.32\textwidth}
        \centering
        \includegraphics[width=\linewidth]{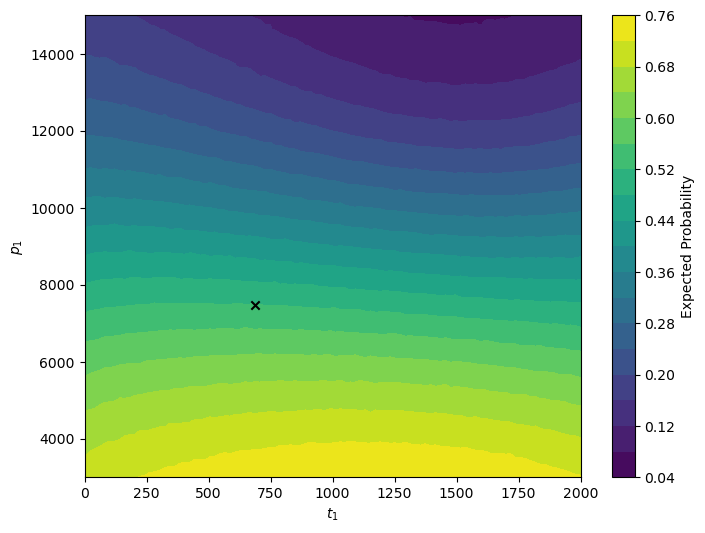}
        \caption{One careful competitor.}
        \label{fig:probs_lateexpensive}
    \end{subfigure}
    \caption{Expected purchase probability against different competitors. Optimal decision marked with \textbf{x}.}
    \label{fig:intelligent_probs}
\end{figure}

   Assume now that the rival companies also exhibit expected utility-maximizing behavior. Thus,
     the first company is modeled as a level-2 adversary as well with respect to the competitors, 
      optimizing its expected utility 
      taking into account forecasts of the 
  competitors' launching behavior drawn on the uncertainty about their expected utility.  For this, we employ Algorithm \ref{alg:generate_sample} to estimate the expected utility of the other companies.
 Next, the competitors' launching decisions are modeled as random variables with probability density functions proportional to their expected utilities, sampled through MC.
We employ BO with a Gaussian process to find the optimal decision $(t_1^*=263, p_1^*=7000)$, resulting in an expected profit of $2,282,981$, smaller than that attained in Section 
 \ref{subsec:example}, since 
  competitors are modeled as strategic and maximizing expected utility. This remarks the 
   importance of modeling in detail 
  competitors' decisions to obtain a more realistic assessment
  of our decisions. The choice of a level-2 model reflects empirical findings in behavioral game theory \citep{camerer2004cognitive, costagomes2006cognition}, which indicate that human and organizational agents rarely exhibit strategic sophistication beyond level-1 or level-2. In competitive business environments marked by private cost functions and asymmetric beliefs, calculating deeper recursive levels introduces spurious precision without empirical support, while sharply compounding computational demands.
  Again, the problem is amenable to brute-force search to further understand the expected utility surface. Figure \ref{fig:utility_intelligent} represents the expected utility with respect to the decision of the advised company, whereas Figure \ref{fig:probs_intelligent} portrays the expected purchase~probability. 

 Importantly, besides modeling the competitors' decisions with uncertainty 
  around their expected utility, we can incorporate prior knowledge about their strategic features. 
  This can stem e.g.\ from market studies, earlier interaction analysis, or qualitative insights regarding their business strategies. For instance, assume we encode whether a competitor is known for prioritizing rapid market entry and aggressive pricing
   through 
  prior distributions, as Appendix A describes. Such assumptions align with real-world marketing strategies, where branding and pre-launch campaigns shape consumer expectations and perceived value. To illustrate this, Figures \ref{fig:utility_cheapearly} and \ref{fig:probs_cheapearly} depict the case of a more aggressive 
   competitor expected to launch earlier and cheaper, potentially capturing a larger market, while the second one behaves in a standard manner. This shifts the optimal decision for the advised company towards longer developing times for a higher quality product, launching the product at a later time.
   
Conversely, Figures \ref{fig:utility_lateexpensive} and \ref{fig:probs_lateexpensive} reflect a scenario where one of the 
 competitors opts for a late but high-quality and expensive launch, possibly due to a commitment to premium product development. 
  There is now just one local optimum for the supported company, as the strategy of delaying development to attain higher quality clearly loses its competitive advantage. Importantly, capturing these nuanced, asymmetric competitor behavior and recalculating the optimal strategic response highlights the primary computational advantage of our framework over rigid, traditional equilibrium models.

\paragraph{Computational scalability}
Regarding scalability with market size, the inner Monte Carlo simulation in Algorithm \ref{alg:generate_sample} requires $\mathcal{O}(M \cdot l)$ operations per objective evaluation. As draws across samples ($M$) and competitors ($l$) are independent, this step parallelizes across $C$ cores as $\mathcal{O}(M \cdot l / C)$. Meanwhile, the outer Bayesian optimization loop scales with cubic cost in the surrogate evaluation budget, decoupled from $M$, $n$, and $l$. Consequently, computation scales linearly with the number of competitors $l$ under vectorized simulation, ensuring practical tractability as the competitive environment expands.

\subsection{Benchmarking against classical frameworks}\label{sec:benchmarking}

To assess the practical benefits of our strategic model, let us benchmark the optimal decision from Section~\ref{kk:sidecars} against three baseline paradigms:
\begin{enumerate}
    \item \emph{Non-adversarial Bayesian model}. Competitors are assumed to act non-strategically according to uniform random distributions, as in the level-0 model of Section~\ref{subsec:example}, and the optimal decision $(t_{\text{naive}}^*, p_{\text{naive}}^*)$ is obtained. 
    \item \emph{Nash equilibrium}. Competitors are assumed to possess full information and common knowledge. Parameter distributions are replaced by their expected values ($\mathbb{E}[a]=0.256$, $\mathbb{E}[c]=0.837$, $\mathbb{E}[\bm{w}]=[0.25, 0.50, 0.25]$, $\mathbb{E}[\rho]=5.0$), and a pure-strategy equilibrium $(t_{\text{NE}}^*, p_{\text{NE}}^*)$ is computed through iterated best response over the discretized grid.
\end{enumerate}
Table~\ref{tab:benchmarks} summarizes the optimal decisions, anticipated profits under each assumed model, and realized performance when each strategy is evaluated against the strategic competitors of Section~\ref{kk:sidecars}, to quantify the loss caused by ignoring competitive intent. 
\begin{table}[htbp]
\centering
\begin{tabular}{l|cc|c|cc|cc}
\hline
Model & $t_1^*$ & $p_1^*$ & $\pi_1$ & Projected & Realized & Loss & Loss \\
 & & & & profit & profit & vs. ARA & (\%) \\
\hline
ARA & 263 & 7000 & 0.449 & 2,282,981 & 2,282,981 & 0 & 0.00\% \\
Naive Bayesian & 283 & 8333 & 0.364 & 2,841,380 & 2,183,766 & 99,214 & 4.35\% \\
Nash equilibrium & 849 & 3000 & 0.672 & 378,150 & 1,267,668 & 1,015,312 & 44.47\% \\
\hline
\end{tabular}
\caption{Comparative performance of strategic decision model against classical baselines.}
\label{tab:benchmarks}
\end{table}
The naive Bayesian policy leads to an overly high price ($p_{naive}^* = 8333$) under the expectation that rivals will price uniformly across $[3000, 15000]$, projecting a profit of $2,841,380$. However, because strategic competitors optimize their own utilities, the firm's realized purchase probability falls to $0.364$, lowering profit to $2,183,766$. This produces an overestimation gap of $657,614$ and a $4.35\%$ reduction in realized earnings relative to the strategic solution.
%
%
Conversely, the Nash equilibrium leads to an aggressive price war down to the boundary $p_{\text{NE}}^* = 3000$ and $t_{\text{NE}}^* \approx 849$ days. When evaluated against competitors who maximize expected utility under private information, this strategy yields $1,267,668$, matching the outcome of the maximin formulation. 

These comparisons show that the proposed approach avoids both the naive overestimation of non-strategic models and the price competition of complete-information 
baselines.

\section{Multiple item purchases}

Consider now this major modeling variation, departing from the initial 
 single purchase assumption per customer, typical of  
  markets with not-so-expensive products and shorter product 
 life-cycles, as would be, e.g., in the case of the video game industry, but clearly not relevant with high-involvement 
 products.
 
 In the incumbent case, each customer will buy as many (different) items as possible to maximize their expected utility taking into account their purchase budget. Given $l$ companies, each with its product with price $p_j$, release time $t_j$ and quality $q_j$, the $k$-th buyer, with budget $b_k$, would solve 
  a knapsack problem~\citep{cacchiani2022knapsack}

\begin{equation}
\begin{aligned}
    & \max_{z_j} \sum_{j=1}^l z_j \cdot u_b(p_j, t_j, q_j) \\
    & {\rm s.t.} \sum_{j=1}^l z_j \cdot p_j \leq b_k, \quad z_j \in \{0, 1\} \quad \forall j,
\end{aligned}
\label{knapsack_eq}
\end{equation}
where $z_j$ indicates whether $(1)$ or not $(0)$ the customer purchases the $j$-th product, providing the portfolio of maximum utility, assuming additive utilities across purchases. 
In particular, the optimal portfolio would include the first company's item if $z_1=1$. 
The discussion extends to cases with generic features $x^j$   as in Section 2, but we adopt these three to extend the case from Section 3.

As before, because of partial information, we would 
model the uncertainty about $u_b$ through a random utility function $U_b$ and about $t_j, p_j, q_j$ through random variables $T_j, P_j, Q_j$ for $j \neq1$, leading to the stochastic 
knapsack problem, 
\begin{equation}
\begin{aligned}
    & \max_{z_j} \,\, z_1 \cdot U_b(p_1, t_1, q_1) + \sum_{j=2}^l z_j \cdot U_b(P_j, T_j, Q_j) \\
    & {\rm s.t.}\,\,\,  z_1 \cdot p_1 + \sum_{j=2}^l z_j \cdot P_j \leq b_k, \quad z_j \in \{0, 1\} \quad \forall j,
\end{aligned}
\label{knapsack_eq2}
\end{equation}
so  that $Prob (z_1=1 | b_k, p_1, t_1 )$  is the probability that the consumer  
buys from the first company, if their budget is $b_k$. Then, the purchase probability would be $\pi (t_1, p_1 ) =
\int Prob (z_1=1 | b, t_1, p_1 ) h (b) \, db $, assuming that $h(b)$ is the budget density in the 
given segment, therefore recovering a mixed binomial purchase model. Existence and continuity 
results are similar to those in Proposition 2, see \cite{lyu2022stochastic,chen2014adaptive} for
different variants and solution approaches.

Algorithm \ref{alg:multiple_items} serves to estimate by MC the expected utility and profit for the first company associated with their decision $(t_1, p_1)$, where we assume a segment of $n$ buyers, each with budget $b_k\sim B, k=1,...,n$ and utility model \eqref{eq:utility}, with uncertainty over its weights and risk aversion coefficient, and, as before, two competitors with the rest of the setup as in Section \ref{subsec:example}. 
  The function $solve\_knapsack$ solves the knapsack problem \eqref{knapsack_eq} for products with utilities $u_j$ and prices $p_j$ for $j=1, 2, 3$, and budget $b_k$, returning the total number of sales of the supported company. 
     Since the number of competitors is small, we use explicit enumeration to solve the knapsack problem. 
Instructions in common with Algorithm 1 are omitted, except 
 where essential. 
This algorithm would feed an expected utility optimization routine as before. 
\begin{algorithm}
\caption{\text{Compute expected utility and profit of decision $t_1, p_1$}} 
\label{alg:multiple_items}
\textbf{Input:} $M$, $n$, $t_1 $, $p_1 $, $c_{11}$, $c_{21}$, $c_{31}$, $e_1(t)$, $B$, $T_j$, $P_j$, $E_j(t)$, $j=2,3$  \\
\textbf{Output:} $util$, $profit$
\begin{algorithmic}[1]
\State $sales = 0$
\State $cost = 0$
\For{$ i \in \{1, \ldots , M \}$}
\State Generate $t_j, p_j, q_j$ for $j=2,3$ as in Alg. \ref{alg:generate_sample} 
\State Generate $e_1, q_1, w_1, w_2 , w_3, \rho$  as in Alg. \ref{alg:generate_sample} 
\State Compute $u_j$ for $ j=1, 2,3$ as in Alg. \ref{alg:generate_sample} 
\State Generate $b_k\sim B, k=1...n$ 
\State $sales = sales + solve\_knapsack(b_k, u_j, p_j)$, $k=1...n$, $j=1,2,3$
\State Compute $c_1 = c_{11} t_1 + c_{21} e_1 + c_{31} q_1$
\State $cost = cost + c_1$
\EndFor
\State $sales = sales / M$
\State $cost = cost / M$

\State $util=u_1(sales \times p_1 - cost)$
\State $profit=sales \times p_1 - cost$
\State \textbf{Return:}  $util$, $profit$
\end{algorithmic}
\end{algorithm} 
 Extensions to cases with multiple buyer segments would follow the path from Section~2.

\begin{figure}[htbp]
    \centering 
    \hfill
    \begin{subfigure}{0.42\textwidth}
        \centering
        \includegraphics[width=\textwidth]{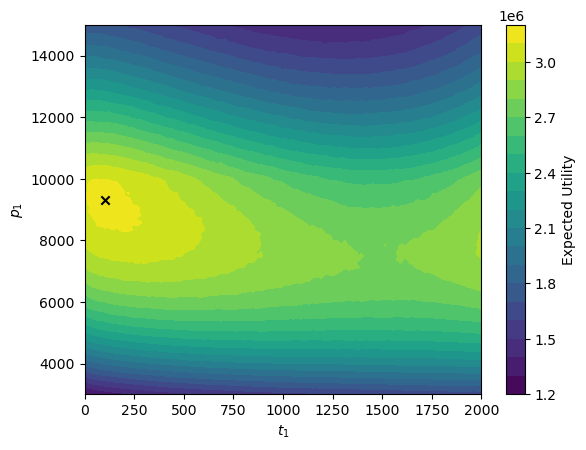}
        \caption{Expected utility.}
        \label{fig:multiple_items_isolines}
    \end{subfigure}
    \hfill
    \begin{subfigure}{0.42\textwidth}
        \centering
        \includegraphics[width=\textwidth]{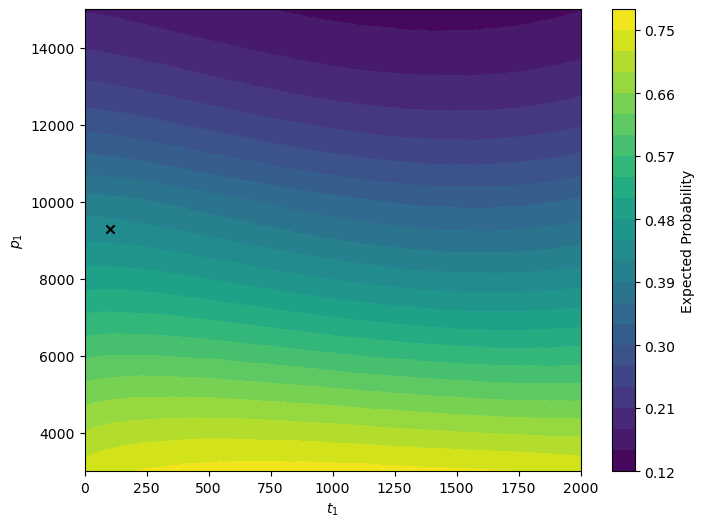}
        \caption{Expected probability of purchase.}
        \label{fig:multiple_prob}
    \end{subfigure}
    \hfill
    \caption{Expected profit and probability with respect to $t_1$ and $p_1$.}
    \label{fig:multiple_items}
\end{figure}

\paragraph{{\bf Experiment}}
  Except when noted, we use the parametric settings from  Section \ref{subsec:example}, as explained in Appendix A. We assume a uniform distribution for the individual budgets $B\sim U[10000, 20000]$
    and perform a brute force search 
 for the optimal launch decision to gain a better understanding of the problem. 
Figure \ref{fig:multiple_items_isolines} represents the expected utility with respect to $t_1$ and $p_1$. The optimal decision would be $(t_1^* =101, p_1^* =9303)$, yielding an expected utility of $3,130,337$. Figure \ref{fig:multiple_prob} represents the expected purchase probability, which is $0.43$ for the optimal decision.
Therefore, it is optimal to launch the product early, 
 since this provides a competitive advantage in a market where a greater utility guarantees sales for a given price. Thus, 
  maximizing the buyer's utility has a bigger role than in Section 3, in which the cost of post-release bugs can shift the decision towards delayed launch times. 

%
\section{Discussion}\label{sec:discussion}

  This paper presents a general structured adversarial risk analysis framework to guide companies in selecting their optimal product launch strategy.
     We emphasized the relevance of handling strategic issues concerning 
     forecasting the behavior of competing companies, as well as those of potential buyers. We adapted ARA-based concepts, stemming from the uncertainty 
     about the competitors' and the consumers' beliefs and preferences, thereby  mitigating common knowledge assumptions
     typical of game-theoretic approaches in the field, including those based on \emph{real options} models.      
      The developed approach facilitates incorporating all relevant uncertain elements to reflect the complexities inherent in real-world market conditions.
       We illustrated the framework with examples recognizing the intricate interplay
   between time, price and quality of a product launch for its intended market. 

Through this research we have contributed towards managerial insights involving the  development and marketing of new niche products, especially with respect to tactical decisions related to `what to launch?', and `when to launch?', important determinants of product success \citep{hultink1997industrial, di1999identifying, calantone2007clustering}. 
Our analysis underscored the importance of early market entry,  when appropriate conditions hold. By entering the market sooner, a company can ensure a foothold, attract early adopters, and build brand recognition and loyalty. It also showed how pricing decisions are interdependent with timing and quality at the time of launch. An appropriate   pricing strategy can enhance market penetration and maximize profits, especially when combined with strategic launch timing. Our examples also showcased how incorporating ARA methods to model strategic buyer and competitor behavior provides more realistic predictions of market responses.  The impact of the number of competitors as well as of risk aversion were 
showcased as well. 

The systematic comparisons in this study provide several actionable principles for product launch management. First, evaluating decisions against strategic rivals demonstrates the tangible cost of strategic naivety. Assuming non-strategic competitor behavior leads the firm to overestimate expected earnings and set excessive prices, incurring market share erosion and profit losses (4.35\% in our benchmark). Second, relying on worst-case maximin models or complete-information Nash equilibria leads to excessive conservatism or destructive price wars, forfeiting substantial profits (up to 44.5\% in our experiments) by driving prices to the lower boundary. Third, the sensitivity analysis uncovers critical tipping points: when post-release defect remediation costs cross a threshold, optimal policy abruptly pivots from aggressive early launch to prolonged testing and higher launch prices. Furthermore, Figure \ref{fig:num_adv} shows that as the number of competitors grows, heightened competitive pressure compels the firm toward later releases and lower prices to defend market share through superior product quality, with steeper utility surfaces reflecting greater strategic sensitivity. Finally, modeling asymmetric competitor profiles reveals that an aggressive, fast-moving rival should be countered with quality differentiation rather than direct price competition, whereas a quality-focused rival requires rapid market entry to capture early market share.

\emph{Empirical calibration and operationalization.}
While the numerical case study relied on historical reliability data \citep{Okumoto1979} combined with stylized strategic parameters to illustrate the mathematical framework, deploying this approach in industry requires calibrating its components from empirical data. Consumer utility parameters and discrete choice weights can be calibrated using choice-based conjoint analysis and revealed-preference transaction logs via hierarchical Bayes estimation. Prior distributions over competitor launch timing, pricing, and quality can be established through structured expert judgment elicitation protocols, supplemented by competitive intelligence from developer releases and patent filings. Furthermore, software reliability growth parameters are readily fitted from historical issue-tracking systems using Bayesian NHPP methods. Once the product is launched, early sales data and competitor announcements could be integrated via sequential Bayesian filtering to dynamically update belief distributions.

\emph{Model boundaries and correlated consumer behavior.}
Our purchasing models assumed that individual buyer decisions are conditionally independent given market choice probabilities. In real-world product diffusion, consumer choices frequently exhibit correlation caused by social contagion, network externalities, and word-of-mouth adoption \citep{bass1969new}. While analytical expressions become intractable under correlated purchasing, our simulation-optimization architecture readily accommodates such dynamics. Specifically, the inner Monte Carlo evaluation can simulate network-based adoption models or Beta-Binomial mixtures to capture correlated market shocks, while the outer Gaussian process surrogate continues to optimize pricing and launch timing as a black-box objective.

Our analysis has been predominantly static, focusing on a single launch event within a product life-cycle. In 
many domains, markets are dynamic, and ongoing post-launch adjustments may be crucial for sustained success. In such dynamic environments it may be relevant to consider reinforcement learning strategies as in \cite{gao2022dynamic}. 

Finally, we assumed that there is only one echelon on the sellers side. However, for many products, such as the launch of a new model of a car by a company, there are several echelons on the seller side, with say the manufacturing car company engaging with several outlets before the buyer can make their decision.  
Extending the framework to multi-echelon scenarios constitute another relevant venue for further research providing a more comprehensive view of product life-cycle and the effectiveness of adaptive strategies.

\section*{Acknowledgments}
Work supported by the Severo Ochoa Excellence Programme CEX-2023-001347-S and the Spanish Ministry of Science program PID2025-172412OB-C21. PGA is a staff member hired under the Generation D initiative, promoted by Red.es, an organization attached to the Ministry for Digital Transformation and the Civil Service, for the attraction and retention of talent through grants and training contracts, financed by the Recovery, Transformation and Resilience Plan through the European Union's Next Generation funds. SD was partially funded by the 2020 Fundaci\'{o}n Mujeres Por África  (Women for Africa Foundation) Award from Spain, which contributed to this collaboration and research. SD also acknowledges the National Institute for Theoretical and Computational Sciences (NITheCS) in South Africa. DRI was supported by the AXA-ICMAT Chair in Adversarial Risk Analysis.


\bibliographystyle{itor}
\bibliography{sample}

\appsection{Appendix A: Parametric choices in the case study}\label{kkappendix}

The appendix provides the specific parameters used to solve the software release case,
 with actual data coming from \cite{Okumoto1979}. Note though that most parametric assumptions were assessed
through sensitivity analysis as detailed below and in
the main text.
 Three of the parameters used come from \cite{softearlier}:
\begin{itemize}
                \item Software life-cycle length $T$. We fix it at $2000$, the time unit being days.
         \item The selling price range for the first company is $p_1 \in [3000, 15000]$.
         \item Assuming risk neutrality, the utility function $u_1$ is, initially, the identity.
    \end{itemize}
The remainder parameters differ from those in \cite{softearlier}. To wit:
    \begin{enumerate}
        \item Cost parameters. For the use case, they are fixed at 
         $c_{11}=0.2 \times 1000$, $c_{21}=1 \times 1000 $, $c_{31}=5 \times 1000 $, where a multiplying factor of $1000$ has been introduced to scale up the cost so it is non-negligible with respect to product price, although these parameters would be adjusted for each specific case. Based on the above, we define the company's cost function $c_1 (t)$ as in (\ref{cost-game}). We perform a sensitivity analysis with respect to $c_{31}$.
        \item Fault discovery process $e_1 (t)$. An NHPP with mean function $m_1 (t)= a t^c$ is employed, obtaining $a$ and $c$ samples from the posterior using Hamiltonian MC implemented with the Python library $pymc$. The prior is a Gamma distribution with mean specified using the parameters in \cite{Zeephongsekul1995}, obtained through MLE based on data available 
         in \cite{Okumoto1979}, leading to estimates $\hat{a}=0.256$ and $\hat{c}=0.837$. The standard deviation is set to $0.1$. Based on this posterior, we compute the required predictions.
    \item When considering level-0 competitors, Section 3.1, these are characterized as follows:
    a) Their release times are modeled as  $T_j \sim U [0,2000], j=2,3$; b) Their prices are modeled as $P_j \sim U[3000,15000], j=2,3$; c) For their random fault discovery process $e_j(t)$ of the competitors, $j=2,3$, we use an NHPP with random mean function
       $m_j (t)= \tilde{a} t^{\tilde{c}}$ with $\tilde{a}\sim Gamma(1.638,0.610)$ and   
       $\tilde{c} \sim Beta \left( 2.019, 0.394 \right)$. Observe that $\mathbb{E}(\tilde{a})=0.256, \mathbb{E}(\tilde{c})=0.837$ and $Var(\tilde{a})=Var(\tilde{c})=0.04$ so that there is more uncertainty about the competitors.
    
    \item In Section 3.2 adversaries are modeled as strategic level-1 thinkers that incorporate prior knowledge.
     They  are characterized as follows:
    a) Their release time is modeled as  $T \sim 2000 \cdot Beta [a,b]$, with $(a=2, b=5)$ for early launching and $(a=5, b=2)$ for late launching; b) Their prices are modeled as $P_j \sim 3000 + 12000 * Beta [a,b]$, with $(a=2, b=5)$ for cheap launching decisions and $(a=5, b=2)$ for expensive ones; c) The random fault discovery process is modeled following the same NHPP as in the previous case. 
    \item Finally we set up the consumer parameters:
    a) The number of potential buyers is $n=1000$; 
     b) A Dirichlet distribution for weights $(w_1,w_2,w_3)$ is $Dir (1,2,1)$, indicating that the weight corresponding to the price of the product is higher due to the greater importance of this factor to buyers, while the other weights remain equally likely; c) A gamma distribution for the customer's risk aversion coefficient $\rho$  is $Ga (5, 1)$, indicating a moderately concentrated distribution around a mean value of 5. We perform sensitivity analysis on this value. 
        \end{enumerate}

\paragraph{Simulated Annealing} We use the following setup for the simulated annealing method. 
Start with a temperature of 1000 and cool it at rate 0.99 over a maximum of 1000 iterations. Neighboring solutions are iteratively generated, evaluating expected utilities, and deciding acceptance based on the temperature and expected utility difference. We generate neighbors by perturbing \( t_1 \) and \( p_1 \) within ranges \([-50, 50]\) and \([-200, 200]\) respectively, and then clipping them to stay within the allowed intervals. 

\appsection{Appendix B: Proofs of Propositions}\label{sec:proofs}

\begin{proof}[Proof of Proposition \ref{lemma_1}]
Let $\bm{x}^1 \in \mathcal{X}^1$, where $\mathcal{X}^1$ is a compact subset of $\mathbb{R}^d$. The expected utility function of the supported company is (1) 
\begin{equation*}
    \psi(\bm{x}^1) = \sum_{j=0}^n \binom{n}{j} \pi(\bm{x}^1)^j (1 - \pi(\bm{x}^1))^{n-j} u_1(j p_1 - c_1(\bm{x}^1)).
\end{equation*}
By hypothesis, the purchasing probability $\pi(\bm{x}^1)$ is continuous in $\bm{x}^1$, which implies that the polynomial factors $\pi(\bm{x}^1)^j (1 - \pi(\bm{x}^1))^{n-j}$ are continuous for each $j \in \{0, \ldots, n\}$. In addition, $u_1(j p_1 - c_1(\bm{x}^1))$ is continuous in $p_1$ and $c_1(\bm{x}^1)$ for each fixed $j$. Because the sum has a finite number of terms, $\psi(\bm{x}^1)$ is a finite sum of products of continuous functions, and is, therefore, continuous on $\mathcal{X}^1$. By the Weierstrass Extreme Value Theorem, any continuous real-valued function defined on a non-empty compact set achieves its supremum, establishing the existence of an optimal decision $\bm{x}^1_* \in \mathcal{X}^1$.
\end{proof}

\begin{proof}[Proof of Proposition 2]
For a given decision $\bm{x}^1$, define the expected utility attained by a buyer from choosing the supported company as (2)
\begin{equation*}
    v_1(\bm{x}^1, u_b, p_b) = \int u_b(\bm{o}(\bm{x}^1), s) p_b(s) ds .
\end{equation*}
  Similarly, for each competitor $i \in \{2, \ldots, l\}$, define $v_i(\bm{X}^i, u_b, p_b) = \int u_b(\bm{o}(\bm{X}^i), s) p_b(s) ds$. Under the assumption that every utility function $u_b$ in the support of $U_b$ is almost surely integrable with respect to $P_b$, the random variables $v_1$ and $v_i$ are finite almost surely. The decision event
\begin{equation*}
    E(\bm{x}^1) = \Big\{ v_1(\bm{x}^1, U_b, P_b) \geq \max_{i \neq 1} v_i(\bm{X}^i, U_b, P_b) \Big\}
\end{equation*}
is measurable in the common probability space, ensuring that $\pi_{ARA}(\bm{x}^1) = \mathbb{P}(E(\bm{x}^1)) = \mathbb{E}[\mathbb{I}_{E(\bm{x}^1)}]$ is well-defined.

To prove continuity, consider a convergent sequence $\bm{x}_k^1 \to \bm{x}^1$. Because $u_b(\bm{o}(\bm{x}^1), s)$ is almost surely continuous in $\bm{x}^1$ and dominated by an integrable function $f(s)$ such that $|u_b(\bm{o}(\bm{x}^1), s)| \leq f(s)$ for all $\bm{x}^1$, the Lebesgue Dominated Convergence Theorem guarantees that
\begin{equation*}
    \lim_{k \to \infty} v_1(\bm{x}_k^1, U_b, P_b) = v_1(\bm{x}^1, U_b, P_b) \quad \text{a.s.}
\end{equation*}
Assuming that the distribution of $v_1(\bm{x}^1, U_b, P_b) - \max_{i \neq 1} v_i(\bm{X}^i, U_b, P_b)$ is atomless so that ties occur with probability zero, the indicator function converges almost surely: $\lim_{k \to \infty} \mathbb{I}_{E(\bm{x}_k^1)} = \mathbb{I}_{E(\bm{x}^1)}$ a.s. Applying the dominated convergence theorem once more yields
\begin{equation*}
    \lim_{k \to \infty} \pi_{ARA}(\bm{x}_k^1) = \lim_{k \to \infty} \mathbb{E}[\mathbb{I}_{E(\bm{x}_k^1)}] = \mathbb{E}[\mathbb{I}_{E(\bm{x}^1)}] = \pi_{ARA}(\bm{x}^1),
\end{equation*}
proving that the choice probability $\pi_{ARA}(\bm{x}^1)$ is continuous in $\bm{x}^1$.
\end{proof}

\appsection{Appendix C: Strategic tipping point in launch timing (Section 3.1)}
\label{sec:appendix_c}

In Section~3.1, expected development and remediation cost under the power-law NHPP discovery curve $m_1(t) = a t^c$ ($0 < c < 1$) is given by:
\begin{equation}
\mathbb{E}[c_1(t_1)] = c_{11} t_1 + c_{31} m_1(T) - (c_{31} - c_{21}) m_1(t_1),
\label{eq:exp_cost}
\end{equation}
with marginal cost derivative:
\begin{equation}
\frac{\partial \mathbb{E}[c_1(t_1)]}{\partial t_1} = c_{11} - (c_{31} - c_{21}) m_1'(t_1) = c_{11} - (c_{31} - c_{21}) a c t_1^{c-1}.
\label{eq:marginal_cost_deriv}
\end{equation}
Because $c < 1$, $m_1'(t_1)$ strictly decreases with $t_1$, meaning bug-fixing savings $(c_{31} - c_{21}) m_1'(t_1)$ diminish over time, making $\mathbb{E}[c_1(t_1)]$ strictly convex in $t_1$.

\begin{proposition}[Threshold for defect cost regime shift]
\label{prop:tipping_point}
Under the NHPP model with concave mean function $m_1(t_1) = a t_1^c$ ($0 < c < 1$):
\begin{enumerate}
    \item[(i)] The derivative \eqref{eq:marginal_cost_deriv} is strictly increasing in $t_1$, so $\mathbb{E}[c_1(t_1)]$ is strictly convex on $(0, T]$.
    \item[(ii)] When cost parameters satisfy
    \begin{equation}
    (c_{31} - c_{21}) a c T^{c-1} > c_{11},
    \label{eq:dominance_condition}
    \end{equation}
    $\frac{\partial \mathbb{E}[c_1(t_1)]}{\partial t_1} < 0$ for all $t_1 \in (0, T]$. Testing strictly reduces expected costs across $[0, T]$, achieving its cost minimum at $t_1 = T$.
    \item[(iii)] Because consumer demand $\pi(t_1, p_1)$ favors earlier launch ($w_1 > 0$), expected profit $\psi(t_1, p_1) = n \pi(t_1, p_1) p_1 - \mathbb{E}[c_1(t_1)]$ is bimodal. There exists a critical threshold $\bar{c}_{31} > c_{21} + \frac{c_{11}}{a c T^{c-1}}$ such that for $c_{31} < \bar{c}_{31}$, the global optimum is an early launch $t_1^* \ll T$, whereas for $c_{31} > \bar{c}_{31}$, the global optimum jumps discontinuously to the boundary $t_1^* = T$.
\end{enumerate}
\end{proposition}

\begin{proof}
For (i), $\frac{\partial^2 \mathbb{E}[c_1(t_1)]}{\partial t_1^2} = (c_{31} - c_{21}) a c (1-c) t_1^{c-2} > 0$ for $t_1 > 0$.
For (ii), since the derivative is strictly increasing, its supremum on $(0, T]$ is $c_{11} - (c_{31} - c_{21}) a c T^{c-1}$, which is strictly negative under \eqref{eq:dominance_condition}.
For (iii), at $t_1 = T$, residual defects vanish ($q_1(T) = 0$), so boundary profit $\psi(T, p_1)$ is independent of $c_{31}$. For any interior $t_1 < T$, residual defects are positive, so $\psi(t_1, p_1; c_{31})$ strictly decreases in $c_{31}$ with slope $-a(T^c - t_1^c) < 0$. If condition \eqref{eq:dominance_condition} fails ($c_{31} \le c_{21} + c_{11}/(acT^{c-1})$), $t_1 = T$ does not minimize costs and early launch strictly dominates. For $c_{31} \to \infty$, interior profit approaches $-\infty$ while boundary profit remains constant. By continuity, there exists a unique threshold $\bar{c}_{31} > c_{21} + c_{11}/(acT^{c-1})$ where the early and boundary peaks coincide, producing the discontinuous jump.
\end{proof}

Evaluating condition \eqref{eq:dominance_condition} with posterior estimates ($\hat{a} = 0.256, \hat{c} = 0.837, T = 2000, c_{11} = 200, c_{21} = 1000$) yields $a c T^{c-1} \approx 0.06208$, requiring $c_{31} > 1000 + 200 / 0.06208 \approx 4222$. Below this threshold, testing to $T$ cannot be optimal; above it, testing reduces costs over the entire window, and once $c_{31}$ crosses $\bar{c}_{31} \approx 4500$--$5000$, cost savings overcome the demand penalty of delay, triggering the regime jump observed in Figure~6b.

\end{document}